\documentclass[aps,graphicx,preprint]{revtex4-2}

\usepackage{hyperref}
\usepackage{amsmath}
\usepackage{array}
\usepackage{dcolumn}
\usepackage{graphicx}
\usepackage{grffile}
\usepackage{siunitx}
\usepackage{mw}
\usepackage{bm}
\usepackage{xcolor}
\usepackage[normalem]{ulem}
\def\Cij{{C_{ij}}}
\def\Wij{{W_{ij}}}
\def\Cklmn{{C_{klmn}}}
\def\Wklmn{{W_{klmn}}}

\begin{document}

\title{Nonlinear deformation and elasticity of BCC refractory metals and alloys}

\author{Vishnu Raghuraman}
\affiliation{Department of Physics, Carnegie Mellon University, Pittsburgh, PA 15213}

\author{Michael C. Gao}
\affiliation{National Energy Technology Laboratory, Albany OR 97321}

\author{Michael Widom}
\affiliation{Department of Physics, Carnegie Mellon University, Pittsburgh, PA 15213}

\begin{abstract}
Application of isotropic pressure or uniaxial strain alters the elastic properties of materials; sufficiently large strains can drive structural transformations. Linear elasticity describes stability against infinitesimal strains, while nonlinear elasticity describes the response to finite deformations. It was previously shown that uniaxial strain along [100] drives refractory metals and alloys towards mechanical instabilities. These include an extensional instability, and a symmetry-breaking orthorhombic distortion caused by a Jahn-Teller-Peierls instability that splays the cubic lattice vectors. Here, we analyze these transitions in depth. Eigenvalues and eigenvectors of the Wallace tensor identify and classify linear instabilities in the presence of strain. We show that both instabilities are discontinuous, leading to discrete jumps in the lattice parameters. We provide physical intuition for the instabilities by analyzing the changes in first principles energy, stress, bond lengths and angles upon application of strain. Electronic band structure calculations show differential occupation of bonding and anti-bonding orbitals, driven by the changing bond lengths and leading to the structural transformations.  Strain thresholds for these instabilities depend on the valence electron count. 
\end{abstract}
\maketitle{}

\section{Introduction}
\label{sec:Intro}
Refractory alloys with high strength and ductility are needed for efficient energy generation and other applications~\cite{rhea1,rhea2,rhea3,rhea4}.  High entropy alloys (HEAs) and other multi-principal element materials (MPEMs) might meet the demands of advanced technologies~\cite{hea1,hea2,hea3,hea4,hea5,miracle}, however optimizing properties and processing in the high dimensional composition space presents a challenge for the design of new alloys~\cite{miracle,yeh}. Fundamental understanding of the elastic and deformation properties and their composition dependence can aid in alloy design~\cite{hea2,miracle,yeh}.

Linear elasticity provides the first indication of the mechanical response of a material to infinitesimal applied stress or deformation. The Voigt tensor $\Cij$, obtained as a second order derivative of free energy with respect to strain, governs the bulk and shear moduli, Poisson ratio and elastic anisotropy \cite{kittel}. The Born stability criteria for elastic stability require that $\Cij$ be positive definite \cite{born}. Large values of the moduli imply high mechanical strength, while high ductility is believed to correlate to some extent with large ratios of bulk modulus to shear modulus~\cite{PughRatio}.

Nonlinear elasticity describes response to finite stress or strain. Under finite stress, the Voigt tensor must be replaced by the Wallace tensor $\Wij$ \cite{wallace,morris}, which still derives from second derivatives of the free energy with respect to strains, but now includes terms arising from the applied stress. Elastic stability requires that $\Wij$ be positive definite \cite{morris}. However, as a differential property, the Wallace tensor only describes infinitesimal deviations from the state of finite strain. It is possible for a structure to be metastable, with a positive-definite Wallace tensor demonstrating local mechanical stability, despite the availability of a lower free energy state upon a certain finite deformation. 

\begin{figure}
	\includegraphics[width=0.5\textwidth]{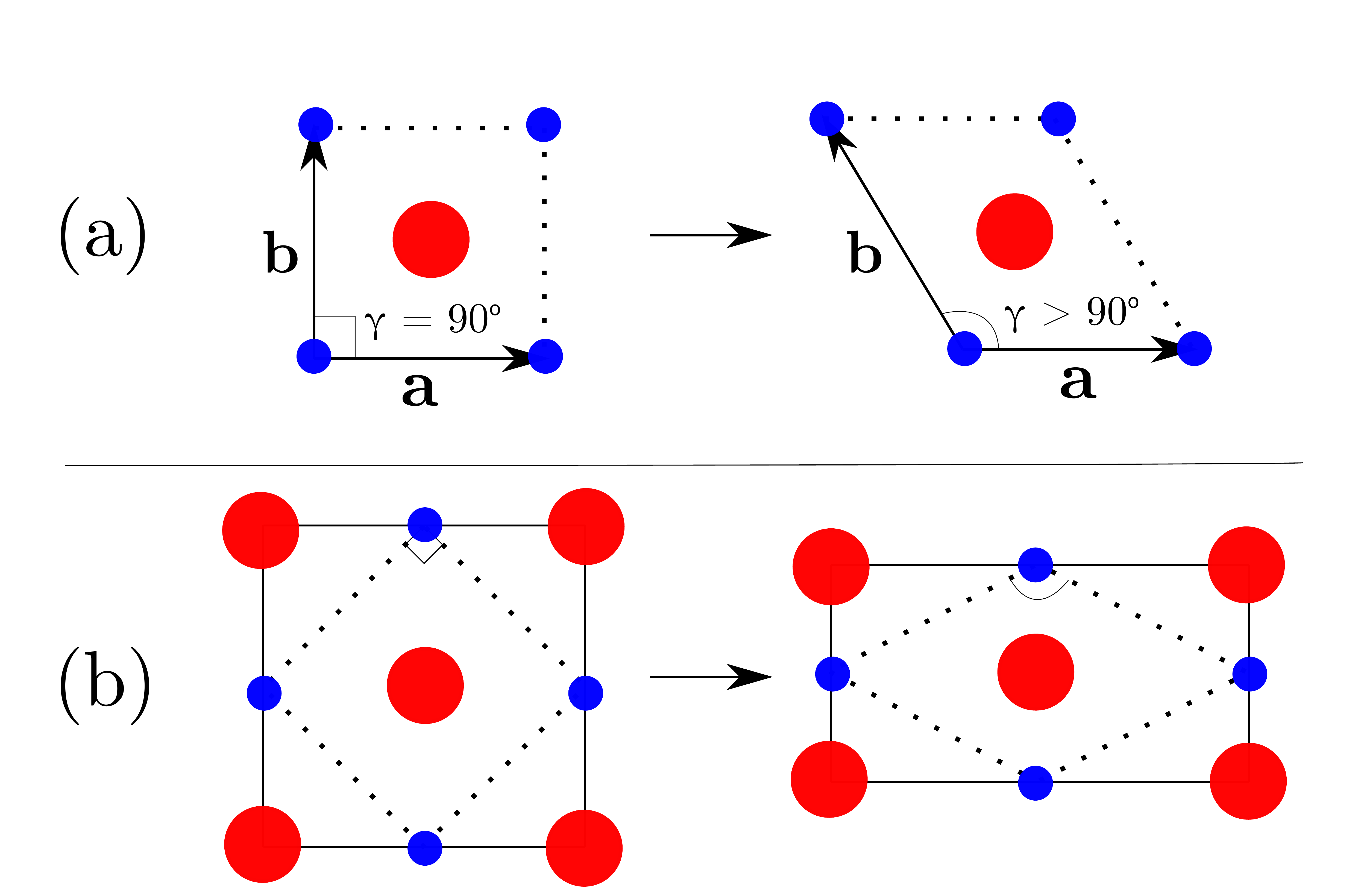}
	\caption{Diagram showing the effect of splay transition on a BCC system seen along [001] direction, represented using (a) 2-atom conventional and (b) 4-atom orthorhombic unit cell. In elemental BCC both atom colors represent the same species at two vertical heights, while in cP2 structures the different colors represent different chemical species.}
	\label{fig:splay-intro-drawing}
\end{figure}
Qi and Chrzan\cite{qi-chrzan} found, based on electronic density functional theory band structure calculations, that many BCC refractory metals exhibit elastic instabilities in which applied uniaxial [100] strain spontaneously breaks the symmetry from body-centered tetragonal to a face-centered orthorhombic structure in which the BCC lattice vectors splay (see Fig.~\ref{fig:splay-intro-drawing}). They assert that the transition is caused by a Jahn-Teller-Peierls distortion \cite{jahnteller,peierls}. Later studies~\cite{dejong, winter} uncovered a second type of transition that we term extensional. This transition is characterized by a sudden jump in $c$-axis and was suggested as a pathway for cleavage. Both transitions are linear elastic instabilities caused by vanishing eigenvalues of the Wallace tensor. The corresponding eigenvectors represent shear in the $xy$-plane at the critical strain $\eta_s$ for the splay transition, while the eigenvector with vanishing eigenvalue has a $zz$-component at the critical strain $\eta_e$ for the extensional instability. An ``intrinsic ductility'' parameter \cite{winter,dejong} $\chi\equiv \eta_e/\eta_s$ was defined to distinguish ductile behavior dominated by shear ($\eta_s<\eta_e$, so that the splay transition would preempt extension) as opposed to brittle behavior ($\eta_e<\eta_s$, so that cleavage occurs). The relationship of $\chi$ to the actual ductility of the material is unclear.

We analyze these transitions in greater detail, with specific attention to alloys of Nb (valence 5) and Mo (valence 6). We elucidate the Jahn-Teller-Peierls mechanism for the splay transition by showing that splay shortens the separation of atoms sharing bonding orbitals, while increasing the separation of atoms connected by anti-bonding orbitals, thus creating a force that drives the transition. The linear instability threshold strain for splay ($\eta_s$) increases with increasing valence electron count. Although the splay occurs in the vicinity of the linear elastic instability, we show that in fact the distortion is discontinuous, caused by a first order transition in the total energy, with hysteresis for strains surrounding the value of $\eta_s$.

\begin{figure}[hbt!]
	\includegraphics[width=0.75\textwidth]{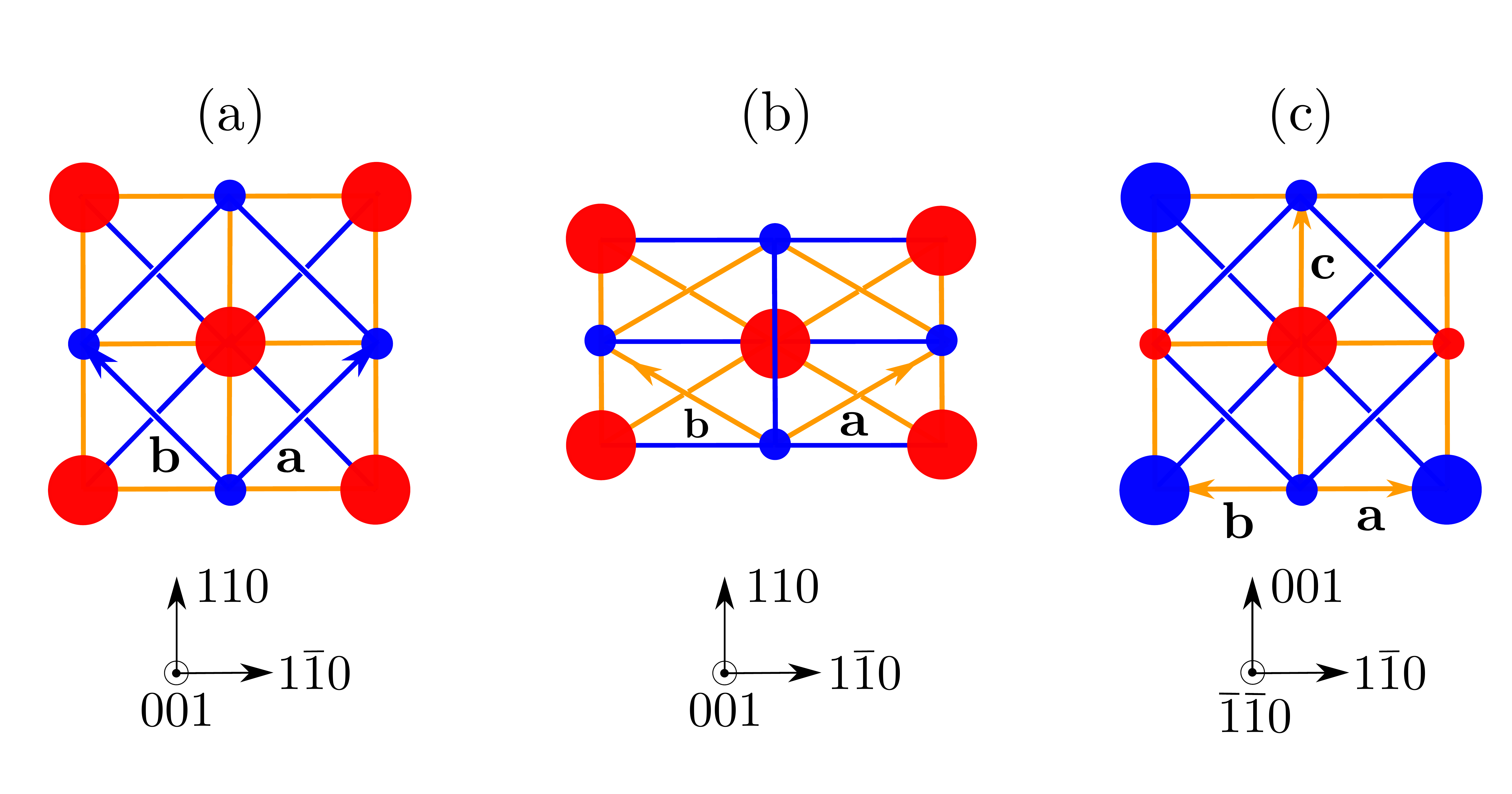}
	\caption{\label{fig:BCC} Alternate views of BCC structures. In elemental BCC both atom colors represent the same species, while in Pearson type cP2 structures the different colors represent different chemical species. Orange bonds are near-neighbors aligned along 3x axes. Blue bonds are next-nearest neighbors aligned along 2x axes. (a) View parallel to [001] with [1$\bar{1}$0] pointing right and [110] pointing up. (b) After stretching along [001] by $c/c_0 = \sqrt{2}$ and splay of $\bm{a}$ and $\bm{b}$ to angle 109.471$^{\circ}$, the BCC structure is restored. Here, as before, the view is along [001] with [1$\bar{1}$0] pointing right and [110] pointing up. Note that the lattice vectors $\bm{a}$ and $\bm{b}$ are now {\em{nearest}} neighbors. (c) View of the stretched and splayed structure along the original [$\bar{1}\bar{1}$0] with [1$\bar{1}$0] pointing to the right and [001] pointing up. Observe that BCC transforms to itself but an initial cP2 structure ceases to be cP2.}
\end{figure}

The extensional instability is also discontinuous, when considered as a function of applied stress rather than strain, with a wide hysteresis region surrounding the stress at which the strain reaches $\eta_e$. When both extension and splay are simultaneously allowed, a curious accident occurs in which the BCC structure reappears in a rotated setting, with the [001] axis turning into [110], and certain near neighbor bonds interchanging with next-nearest neighbors (see Fig \ref{fig:BCC}).

In the following we describe our calculation procedures. We then recreate the Qi and Chrzan\cite{qi-chrzan} result for the case of elemental Nb, analyze the geometry of the structure at various strains along with the nature of the transition, and link them with electronic band structure calculations. The symmetry-breaking is also explained by analyzing the eigenvalues and eigenvectors of the symmetrized Wallace tensor. Implications of the Wallace tensor calculations on the ductility are discussed.  Then we vary the VEC, first increasing from VEC=5 to 6 by alloying Nb with Mo, then reducing to VEC=4.5 at composition NbZr. The effect of this variation on the symmetry-breaking transition is explored. Finally we discuss the implications for design of HEAs and MPEMs.

\section{Methods}
\label{sec:Methods}

Our calculations are based on electronic density functional theory band structures and total energies as obtained using the program VASP \cite{vasp}. We apply PAW potentials (Mo\_pv 2005, Nb\_pv 2002, Zr\_sv 2005) in the PBE generalized gradient approximation \cite{gga}. An enhanced plane wave basis energy cutoff of 340 eV is used, and Methfessel-Paxton first order smearing \cite{ismear1} of width 0.2 eV. Our $k$-point grids densities exceed 30,000 $k$-points per reciprocal atom (25x25x25 grid for a 2-atom cubic cell, 20x20x20 grid for a 4-atom cubic cell and 13x13x13 for a 16-atom cubic cell). Full resolution Fourier transform grids are obtained using ``Accurate'' precision settings.

Electronic band structures are obtained from VASP, along with projections of states onto the atomic $s$- and $d$-orbitals. To capture the effect of the splay transition, we follow high symmetry paths through the Brillouin zone for crystals with face-centered orthorhombic symmetry. Wave functions at special points are plotted using the program {\tt WaveTransPlot2D}~\cite{wavetrans}.

All structures are fully relaxed subject to certain constraints that we specify such as symmetry or uniaxial strain. In order to obtain the true stable state, in the presence of possible symmetry-breaking, we begin our relaxations from a state consistent with the symmetry of the stable state. Uniaxial strains are maintained by modifying the VASP code to zero out the $c$-axis components of stresses. For comparison, we also relax the set of uniaxially strained structures while maintaining tetragonal symmetry.

The second order elastic tensors $\Cklmn$ are obtained by calculating stresses $\sigma_{kl}$ in response to applied strains $u_{mn}$ (here $k$, $l$, $m$, $n$ range over the Cartesian indices $x, y, z$). The resulting matrix is averaged with its transpose to impose symmetry. Consider a structure that has been deformed by a Green-Lagrange strain $\bm{\eta}$. This system is in mechanical equilibrium if, on application of additional infinitesimal strain $\delta u(\bm{x})$, the free energy change
\begin{equation}
	\delta F[\delta u(\bm{x})] \geq 0. 
	\label{eq:free-energy}
\end{equation}
With significant algebra\cite{morris}, (\ref{eq:free-energy}) can be rewritten as
\begin{equation}
	W_{klmn}\delta u_{kl}\delta u_{mn} \geq 0.
\end{equation}
Here $W_{klmn}$ is the symmetrized Wallace tensor, given by \cite{wallace,morris}
\begin{equation}
	\label{eq:Wallace}
	\Wklmn = C^{\prime}_{klmn} + \frac{1}{2}\left[\tau_{ml}\delta_{kn} + \tau_{km}\delta_{ln} + \tau_{nl}\delta_{km} + \tau_{kn}\delta_{lm} - \tau_{kl}\delta_{mn} - \tau_{mn}\delta_{kl}\right],
\end{equation}
where $\bm{\tau}$ is the second Piola-Kirchoff stress. It is more convenient to use Voigt notation for the elastic and Wallace tensors. In this notation, $\Cklmn \xrightarrow[]{} \Cij$, $\Wklmn \xrightarrow[]{} \Wij$, where $1 \leq i, j \leq 6$, with $1 \xrightarrow[]{} xx,\;2 \xrightarrow[]{} yy,\;3 \xrightarrow[]{} zz,\;4 \xrightarrow[]{} yz,\;5 \xrightarrow[]{} xz,\;6 \xrightarrow[]{} xy$. The eigenvectors of $\Wij$ fall into two types: extensional, with non-vanishing projections along the $zz$ direction; and shear, with only $xy, yz, zx$ components. If all eigenvalues are positive, the structure is mechanically stable against infinitesimal distortions, while a vanishing eigenvalue indicates the onset of a linear elastic instability. We call the instability extensional if it has a $zz$ component, and we call it shear otherwise. For further information on the derivation of the Wallace tensor, the reader is referred to the work of Morris and Krenn\cite{morris}. For initially cubic systems that have been stretched along the [001] direction, the tetragonally symmetric Wallace tensor is given by
\begin{equation}
	W_{ij} = \begin{pmatrix}
		C^{\prime}_{11} & C^{\prime}_{12} & C^{\prime}_{13} - \frac{\tau}{2} & 0 & 0 & 0 \\
		C^{\prime}_{12} & C^{\prime}_{11} & C^{\prime}_{13} - \frac{\tau}{2} & 0 & 0 & 0 \\
		C^{\prime}_{13} - \frac{\tau}{2} & C^{\prime}_{13} - \frac{\tau}{2} & C^{\prime}_{33} + \tau & 0 & 0 & 0 \\
		0 & 0 & 0 & C^{\prime}_{44} + \frac{\tau}{2} & 0 & 0 \\
		0 & 0 & 0 & 0 & C^{\prime}_{44} + \frac{\tau}{2} & 0 \\
		0 & 0 & 0 & 0 & 0 & C^{\prime}_{66}
	\end{pmatrix},
\end{equation}
where $\tau$ is the second Piola-Kirchoff stress along the 33 direction. However, once the tetragonal symmetry breaks, the form of the tensor changes to orthorhombic, 
\begin{equation}
	W_{ij} = \begin{pmatrix}
		C^{\prime}_{11} & C^{\prime}_{12} & C^{\prime}_{13} - \frac{\tau}{2} & 0 & 0 & 0 \\
		C^{\prime}_{12} & C^{\prime}_{22} & C^{\prime}_{23} - \frac{\tau}{2} & 0 & 0 & 0\\
		C^{\prime}_{13} - \frac{\tau}{2} & C^{\prime}_{23} - \frac{\tau}{2} & C^{\prime}_{33} + \tau & 0 & 0 & 0 \\
		0 & 0 & 0 & C^{\prime}_{44} + \frac{\tau}{2} & 0 & 0 \\
		0 & 0 & 0 & 0 & C^{\prime}_{55} + \frac{\tau}{2} & 0 \\
		0 & 0 & 0 & 0 & 0 & C^{\prime}_{66}
	\end{pmatrix}.
\end{equation}
Appendix \ref{app:A} derives the eigenvalues and eigenvectors for the symmetrized Wallace tensors.

A related approach was taken by de Jong\cite{dejong} and by Winter\cite{winter}, who evaluated the Wallace tensor perturbatively, using second- and third-order elastic constants evaluated at zero strain in order to estimate the Wallace tensor at small strains. Our approach is more accurate at large strain, as it evaluates the actual tensor in the presence of strain. The prior study termed an instability {\em cleavage} if the eigenvector contained a $zz$ component. Here we prefer the term {\em extensional} because it is not certain that the material will break in response to the instability, and cleavage may arise from other causes.

\section{Elemental N\lowercase{b} and M\lowercase{o}}

\label{sec:Nb-Mo}

\subsection{Energetics}
Let's start from the fully relaxed body centered cubic state (Pearson type cI2, strukturbericht A2) in a 4-atom unit cell as shown in Figure \ref{fig:BCC}a . The splay angle $\gamma$, which is defined as the angle between $\ba$ and $\bb$, is 90$^{\circ}$. The dominant slip system in BCC metals is \{110\}$<$111$>$ (i.e. the closest packed plane and the shortest slip distance). High-order slip planes in BCC metals include \{112\}, \{123\}, and others \cite{slip1,slip2} . Previous ab-initio uniaxial tension tests on BCC W \cite{MoWslip1} and BCC Mo \cite{MoWslip2} show that the maximum tensile stress is the lowest along $<$100$>$ direction compared with $<$110$>$ and $<$111$>$ directions. We apply a uniaxial extension of the $c$-axis ([001]) by fractional values ranging from -10\% up to +70\%, yielding body-centered tetragonal structures (Pearson type tI2, Strukturbericht A$_a$). The deformed structure is then relaxed subject to the constraint of constant $zz$ strain. Separately, we apply additional pure shear in the $xy$ plane to transform to a face-centered orthorhombic structure (Pearson type oC4, Strukturbericht A20) and then relax at constant $zz$ strain.

\begin{figure}
	\includegraphics[width=0.5\textwidth]{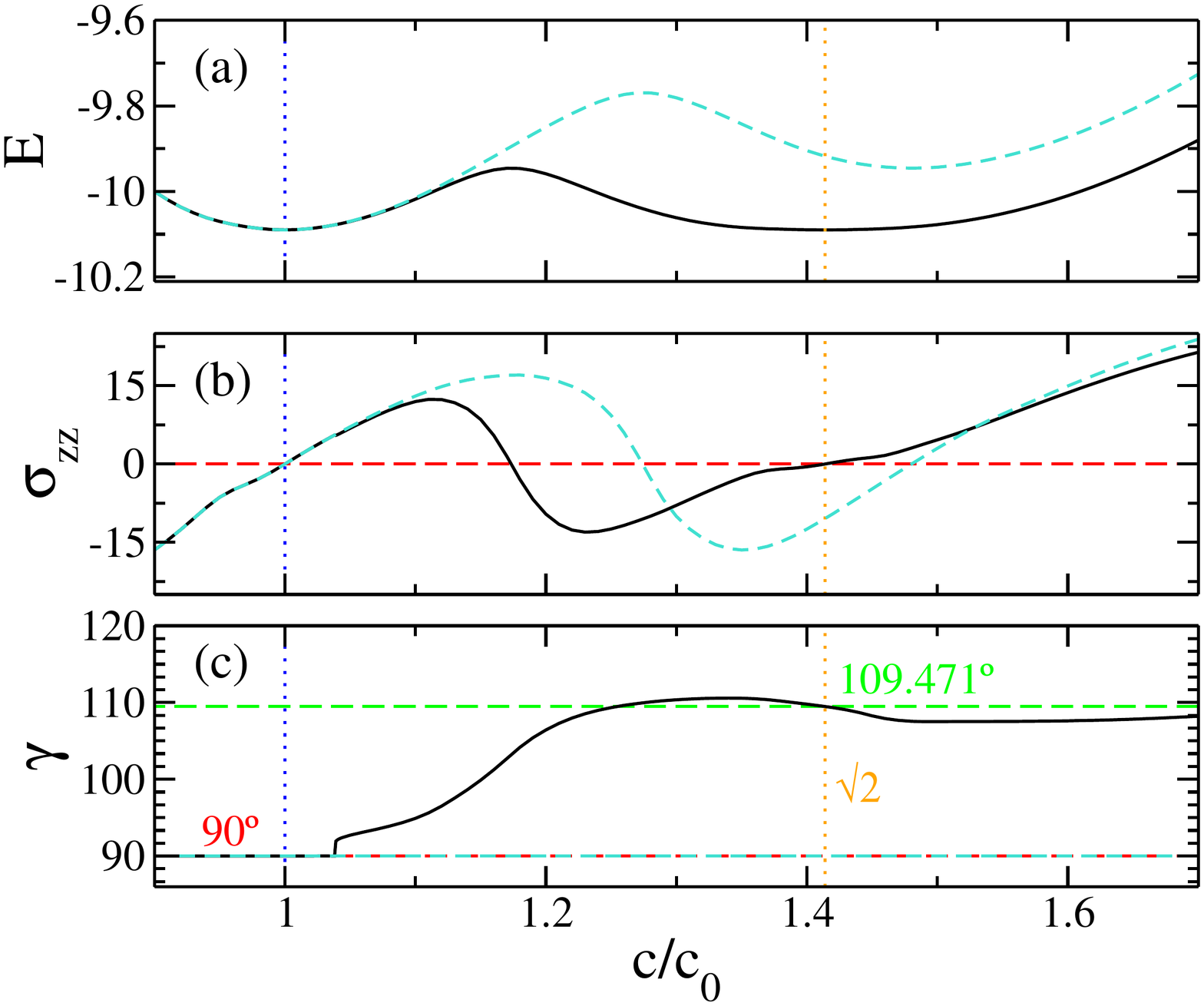}
	\caption{\label{fig:Nb-data} Data for elemental Nb under uniaxial strain with constant $c$ ($c_0$ is the equilibrium value). Graphs (a), (b), and (c), show energy $E$ (units eV/atom), stress $\sigma_{zz}$ (units GPa), and unit cell angle $\gamma$ (units $^\circ$), respectively. Black curves arise from initial orthorhombic distortion while dashed turquoise apply to tetragonal symmetry.}
\end{figure}

\begin{figure}
  \includegraphics[width=0.5\textwidth]{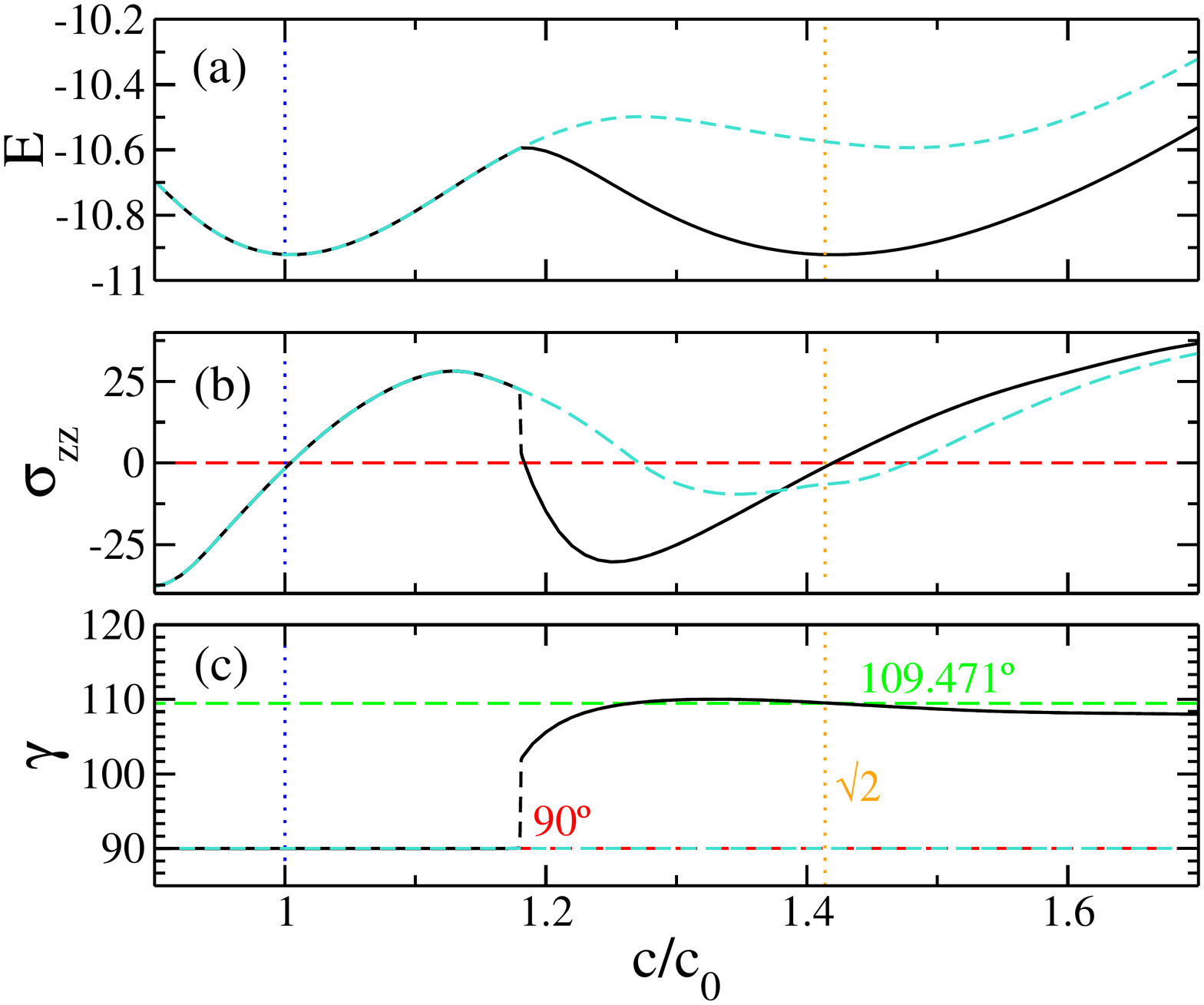}
  \caption{\label{fig:Mo-data} Data for elemental Mo under uniaxial strain with constant $c$ ($c_0$ is the equilibrium value). Graphs (a), (b), and (c), show energy $E$ (units eV/atom), stress $\sigma_{zz}$ (units GPa), and unit cell angle $\gamma$ (units $^\circ$), respectively. Black curves arise from initial orthorhombic distortion while dashed turquoise apply to tetragonal symmetry.}
\end{figure}

Figures~\ref{fig:Nb-data} and \ref{fig:Mo-data} illustrate our results for pure Nb and Mo respectively. Let's first look at Nb. Part (c) shows that the structure relaxes to tetragonal up to approximately 4\%, after which the splay angle $\gamma$ grows and the structure becomes orthorhombic. With further stretching, the splay angle seems to stabilize around the value 109.471$^\circ$ ($\arccos(-1/3)$), which is the angle between BCC primitive vectors. Additionally, at 41.4\% strain ($c/c_0 = \sqrt{2}$), the energy is identical to the undistorted BCC structure. It turns out that our distorted orthorhombic structure has reverted to a two-atom supercell of the BCC primitive cell, oriented so that the conventional BCC [110] axis lies parallel to the Cartesian $z$ axis. Fig.~\ref{fig:BCC} illustrates the distortion sequence. Start with a conventional BCC cell, with the $\ba$, $\bb$, $\bc$ vectors aligned with the [001], [010], and [001] directions, respectively as illustrated in Fig.~\ref{fig:BCC}a. After stretching the $\bc$ vector $\sqrt{2}$ along [001] direction, and relaxing, the $\ba$ and $\bb$ vectors shorten and their angle grows from 90$^\circ$ to 109.471$^\circ$, as seen in Fig.~\ref{fig:BCC}b. Viewing in a perpendicular direction, as in Fig.~\ref{fig:BCC}c, reveals that the combined stretching and splaying carries BCC structures back to themselves.

Notice in Fig. \ref{fig:Nb-data}b the maximum of stress happens at around 12\% strain. Further stretching results in the decrease of stress, which implies a mechanical instability \cite{born}. If tensile stress is kept fixed, the c-axis will jump discontinuously from 12\% to nearly 60\%. A discontinuous jump will also occur from 22\% to -8\% on compression. In this case the system will also revert from orthorhombic ($\gamma > 90^{\circ}$) to a tetragonal structure with $\gamma = 90^{\circ}$. It is also evident that the orthorhombic distortion reduces the stress relative to what it would have been if tetragonal symmetry were maintained. This stress relaxation may enhance the ductility of Nb.

Figure \ref{fig:Mo-data} shows a similar story for Mo. In this case, the splay angle jumps to 102$^\circ$ at around 18\% strain. However, the stress maximum lies at 13\% strain, implying that the orthorhombic transition occurs after the system becomes extensionally unstable. The implications of this switched order will be apparent when we discuss the Wallace tensor eigenvalues.

\subsection{Elasticity}
\label{sec:elastic}

\begin{figure}
	\centering
	\includegraphics[width=0.5\textwidth]{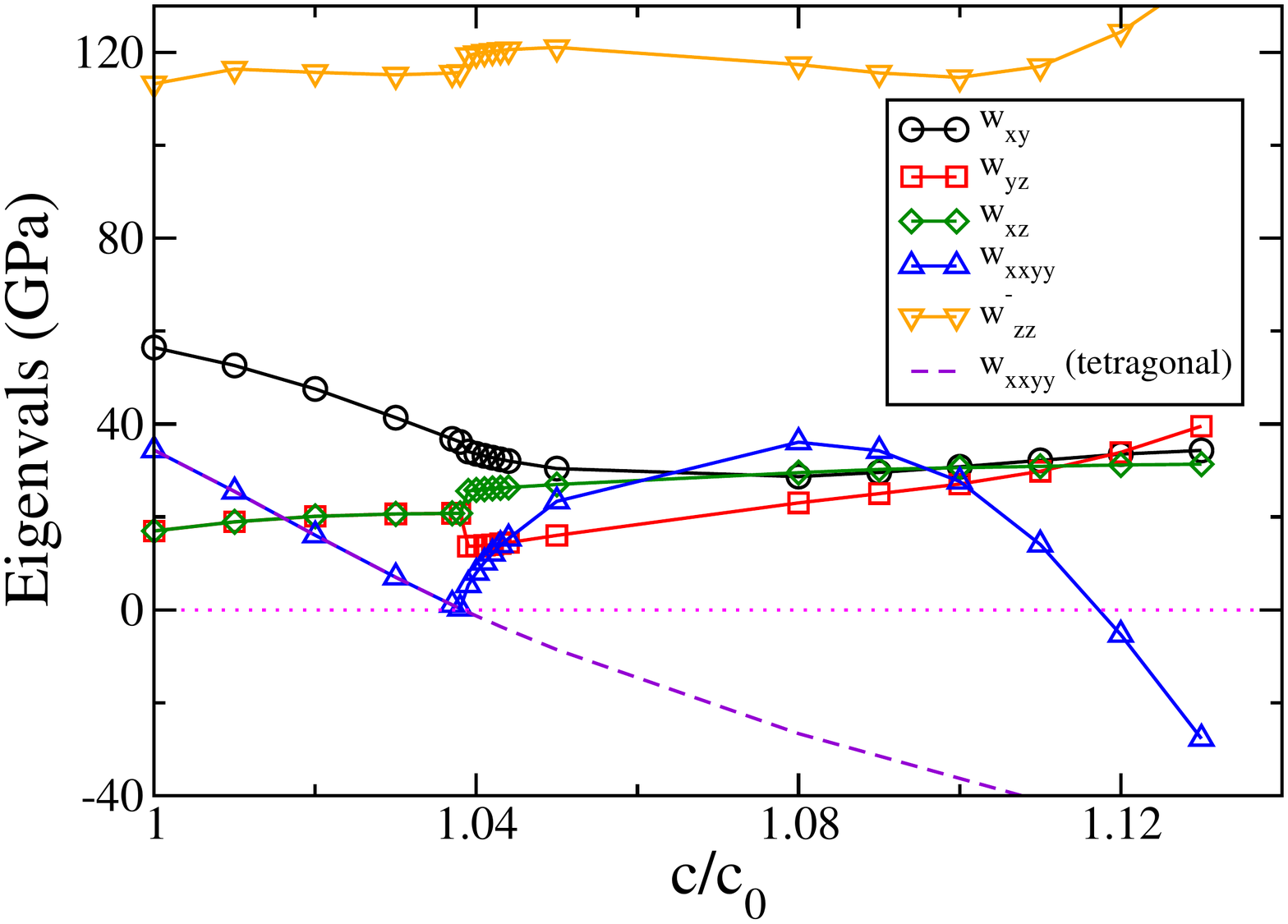}
	\caption{The eigenvalues of the symmetric Wallace Tensor for Nb as a function of the uniaxial strain. Eigenvalue $w^{+}_{zz}$ lies off-scale at around 500 GPa.} 
		\label{fig:swt-Nb}
\end{figure}

Let's analyze the evolution of the Wallace tensor eigenvalues for uniaxially strained Nb, shown in Figure \ref{fig:swt-Nb}. The scripts used to calculate the symmetric Wallace tensor eigenvalues are made available on the internet\cite{github}. Up to (and including) 3.8\% strain, the tetragonal system is used to calculate eigenvalues, and from 3.9\% strain onward the orthorhombic solution is used.  The eigenvalues are labeled $w_{xy},w_{xz},w_{yz},w_{xxyy}, w^{+}_{zz}$ and $w^{-}_{zz}$, and the corresponding eigenvectors are labelled $v_{xy},v_{xz},v_{yz},v_{xxyy}, v^{+}_{zz}$ and $v^{-}_{zz}$.  The analytical expressions of the eigenvectors, available in Appendix \ref{app:A}, determine the mode of instability.  For the tetragonal structure, $v_{xy},v_{xz},v_{yz},v_{xxyy}$ represent shear modes while $v^{+}_{zz}$ and $v^{-}_{zz}$ represent extensional modes. 

Two main ``signatures" of the orthorhombic transition can be seen in the eigenvalue plot. Firstly, the degeneracy of the $v_{xz}$ and $v_{yz}$ eigenvalues is split above 3.8\% strain due to the loss of tetragonal symmetry. Secondly, the behavior of the $v_{xxyy}$ eigenvalue changes near the transition. The dashed line denotes the trajectory of the $v_{xxyy}$ eigenvalue when tetragonal symmetry is constantly enforced. The tetragonal eigenvalue goes negative, while the orthorhombic eigenvalue increases at the transition and remains positive. A negative eigenvalue means that the system is mechanically unstable \cite{morris,dejong,winter}. The incipient vanishing of $w_{xxyy}$ creates transverse phonon softening and could lead to a domain structure. Hence, we can say that the orthorhombic transition enables the uniaxially strained Nb system to remain mechanically stable, and acts as a mechanism to boost intrinsic ductility by relieving stress.

\begin{figure}
	\centering
	\includegraphics[width=0.5\textwidth]{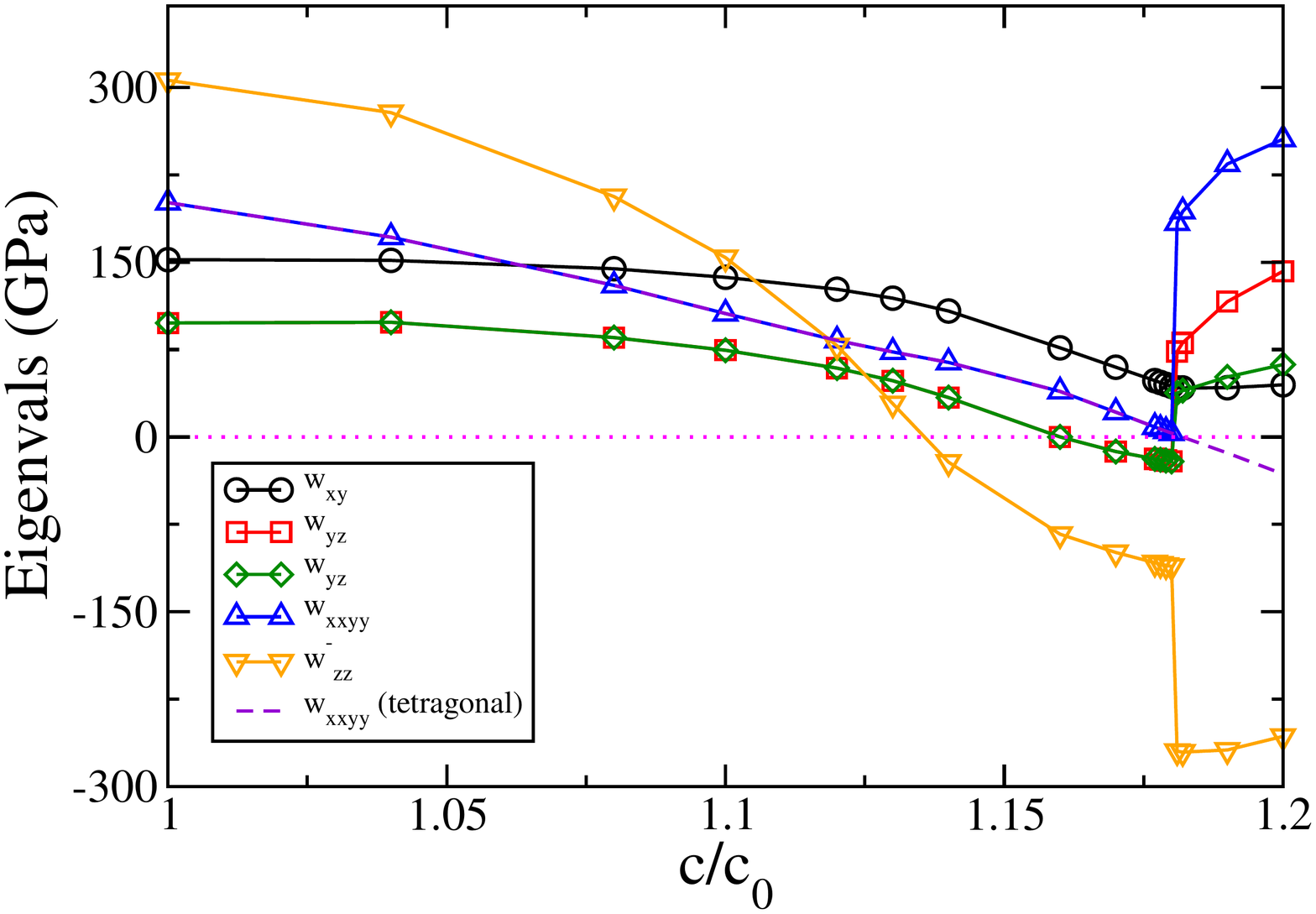}
	\caption{The eigenvalues of the symmetrized Wallace tensor for Mo as a function of the uniaxial strain. Eigenvalue $w^{+}_{zz}$ lies off-scale at around 750 GPa}
	\label{fig:Mo-swt}
\end{figure}

Figures \ref{fig:Mo-swt} contain similar plots for the Mo system. While the $v_{xxyy}$ eigenvalue behaves the same way as observed in Nb, the orthorhombic transition takes place {\em{after}} the extensional mode ($v^{-}_{zz}$) has become negative, implying that the material ``breaks" extensionally before the orthorhombic distortion occurs.  This conclusion agrees with the observation in Section \ref{sec:Nb-Mo}A, where the peak of the tensile stress occurred before the jump in the splay angle.  As a result, Mo jumps discontinuously from tetragonal to an orthorhombic structure. The orthorhombic instability is hidden inside the extensional instability.

\section{Orthorhombic Transition - Geometric Details}
\label{sec:geometry}
\subsection{Contour plots}
\begin{figure}
    \centering
    \includegraphics[width=0.9\textwidth]{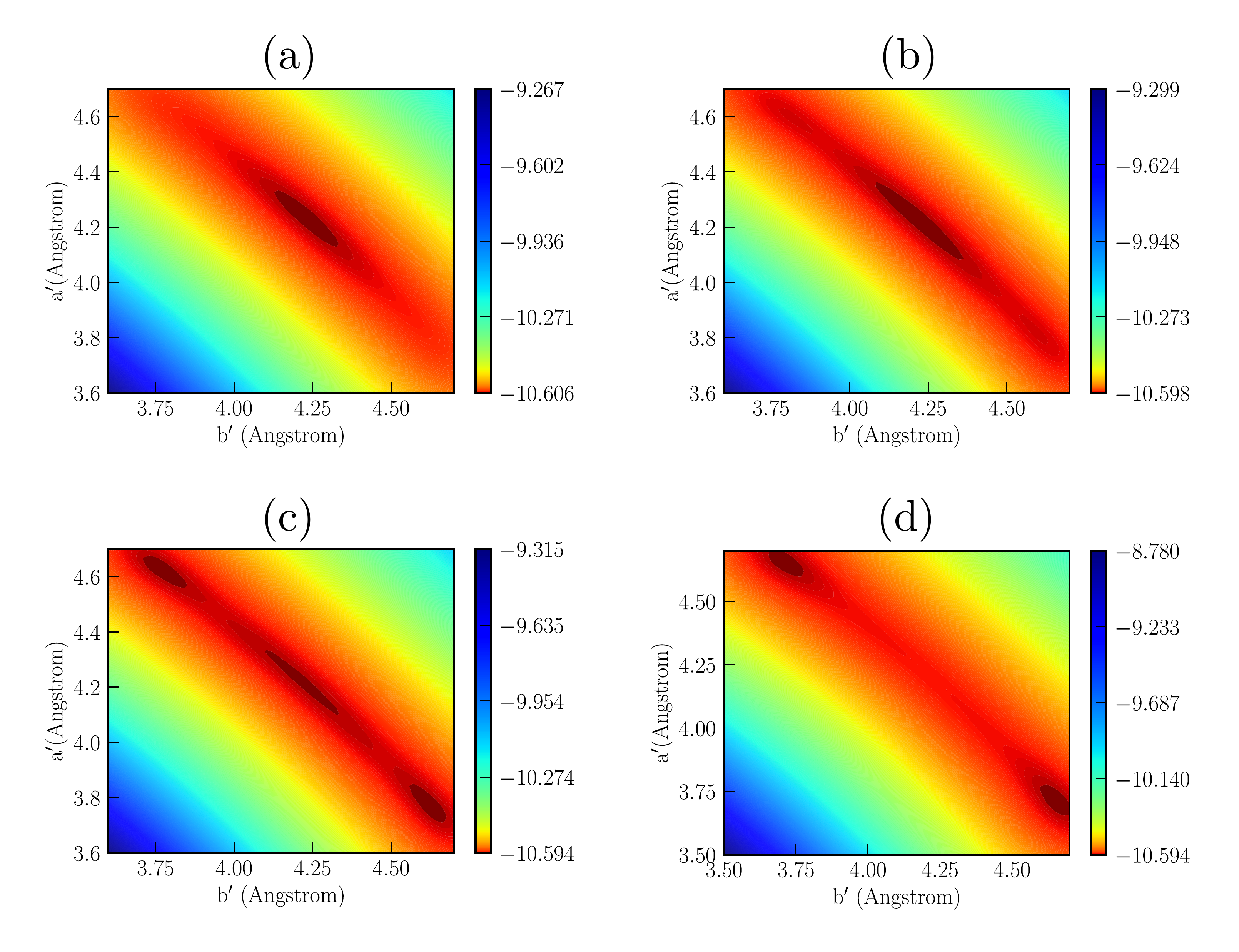}
    \caption{Contour plot of energy per Mo atom (in eV) as functions of the orthorhombic lattice constants $a^{\prime}$ and $b^{\prime}$ (in \AA) at (a) 17.5\% strain along $z$-direction; (b) 17.9\% strain; (c) 18.1\% strain; (d) 18.5\% strain.}
    \label{fig:Mo-17.5}
  \end{figure}

Notice the discontinuous jumps in $\gamma$ vs $c/c_0$ shown for Nb in Figure \ref{fig:Nb-data}c and for Mo in Figure \ref{fig:Mo-data}c. The tetragonal to orthorhombic transitions are discontinuous, and possess small hysteresis regions. To see this, we demonstrate the simultaneous existence of two locally stable states whose energies interchange with increasing applied strain. Let $a^\prime$, $b^\prime$, and $c$ be the lattice parameters of an orthorhombic structure. Figure~\ref{fig:Mo-17.5} graphs the energy of Mo as a function of $a^\prime$ and $b^\prime$ for values of $c$ passing through the transition. At applied strain of 17.5\% (part (a)) a single energy minimum is visible, with $a^\prime = b^\prime$ implying the structure is tetragonal. At 17.9\% a pair of additional local minima appear with $a^\prime\neq b^\prime$, but their energies exceed the tetragonal case. Beyond 18.1\% the orthorhombic energy drops below the tetragonal energy, and after 18.5\%, the tetragonal minimum vanishes. The presence of a region with multiple co-existing phases separating the two single phase regimes (tetragonal and orthorhombic) is characteristic of discontinuous transitions. To understand why this transition is discontinuous, the reader is directed to Appendix \ref{app:B}, which contains a simple model that produces such a phase transition. The model also produces quantitative results for the Nb transition which compare well to first-principles data.

\subsection{Bond length dimerization}
\begin{figure}
    \centering
    \includegraphics[width=0.6\textwidth]{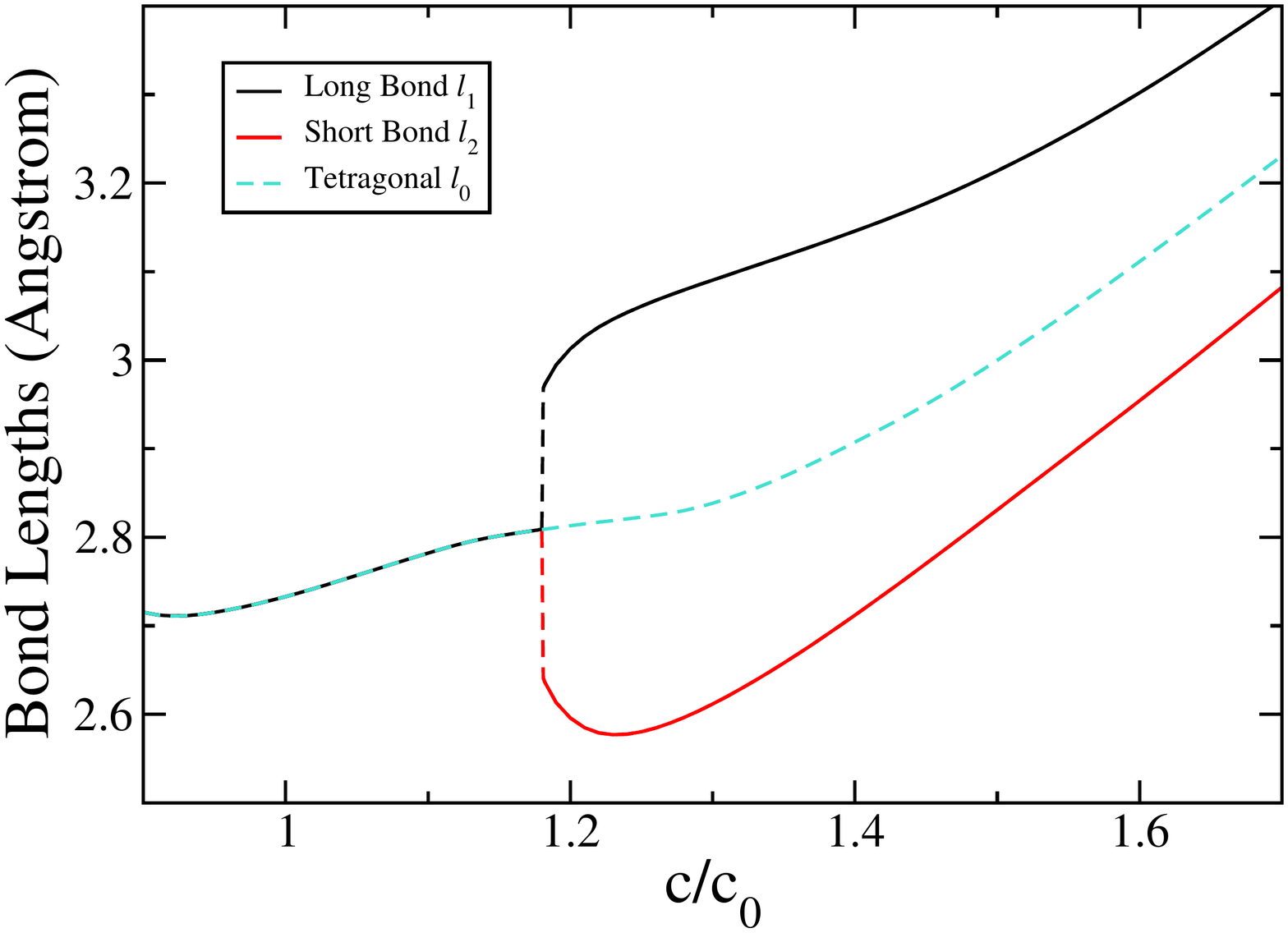}
    \caption{The evolution of the long ($\bm{n}_1 = \frac{1}{2}\left(\bm{a} - \bm {b} + \bm{c}\right)$) and the short ($\bm{n}_2 = \frac{1}{2}\left(\bm{a} + \bm{b} + \bm{c}\right)$) bonds with the applied strain for Mo. The blue dashed line represents the near neighbor bond length for tI2 Mo.}
    \label{fig:bond-strain-evolution}
\end{figure}

\begin{figure}
    \centering
    \includegraphics[width=0.5\textwidth]{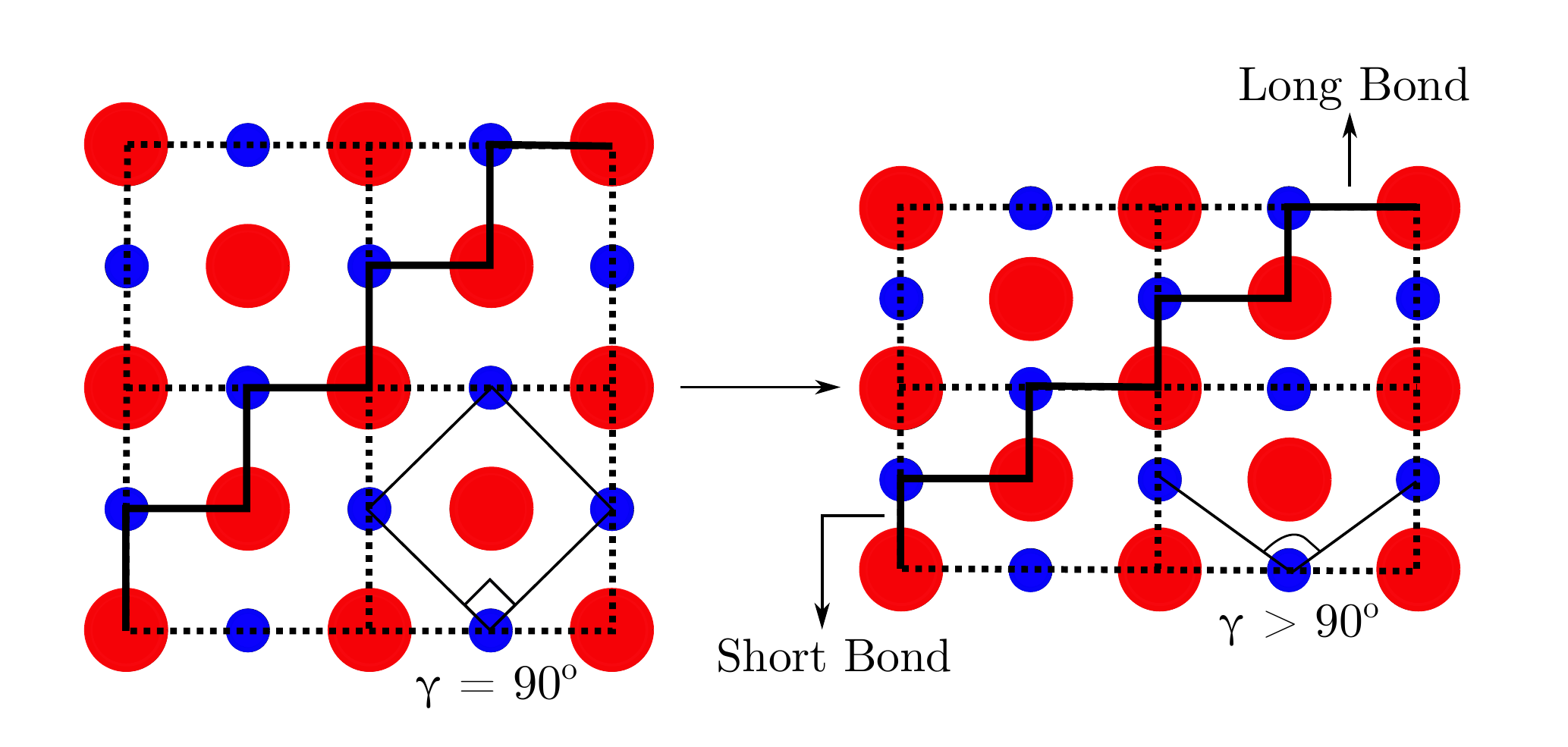}
    \caption{Pictorial depiction of the tetragonal to orthorhombic transition as a Peierls distortion, seen along the [001] direction. Large (red) and small (blue) atoms are at two different vertical heights. The figure on the left is tetragonal Mo at 10\% strain and the figure on the right is orthorhombic Mo at 30\% strain, well beyond the transition. Dotted lines have been used to highlight the cells, and the bold solid line denotes the near-neighbor bond chain. The thin solid lines represent two-atom cells.}
    \label{fig:peierls}
\end{figure}

Consider the near neighbor bond vectors $\bm{n}_1 = \frac{1}{2}(\ba - \bb + \bc)$ and $\bm{n}_2 = \frac{1}{2}(\ba + \bb + \bc)$. In terms of orthorhombic lattice constants, these bonds have lengths $\ell_1=\frac{1}{2}\sqrt{a^{\prime 2} + c^2}$ and $\ell_2=\frac{1}{2}\sqrt{b^{\prime 2} + c^2}$ respectively. Figure \ref{fig:bond-strain-evolution} traces the evolution of these bond lengths with applied strain. Before the transition $\ell_1=\ell_2 = \ell_0 =\frac{1}{2} \sqrt{a^2 + c^2}$, where $a$ and $c$ are the tetragonal lattice constants. At 18.1\% strain the bond length curves bifurcate, resulting in orthorhombic distortion, with (arbitrarily) $b^\prime<a^\prime$, and hence $\ell_2<\ell_1$. We refer to $\bm{n}_1$ as the ``long bond'' and $\bm{n}_2$ as the ``short bond'', even though both bonds were nearest-neighbors prior to the transition. Symmetry breaking leading to an alternation of bond lengths (see Fig.~\ref{fig:peierls}), also known as dimerization, is a characteristic feature of the Peierls transformation~\cite{peierls}.

\section{Electronic Structure Analysis of the Transition}
\label{sec:bands}
\subsection{Nb}
\begin{figure}
    \centering
    \includegraphics[width=0.49\textwidth]{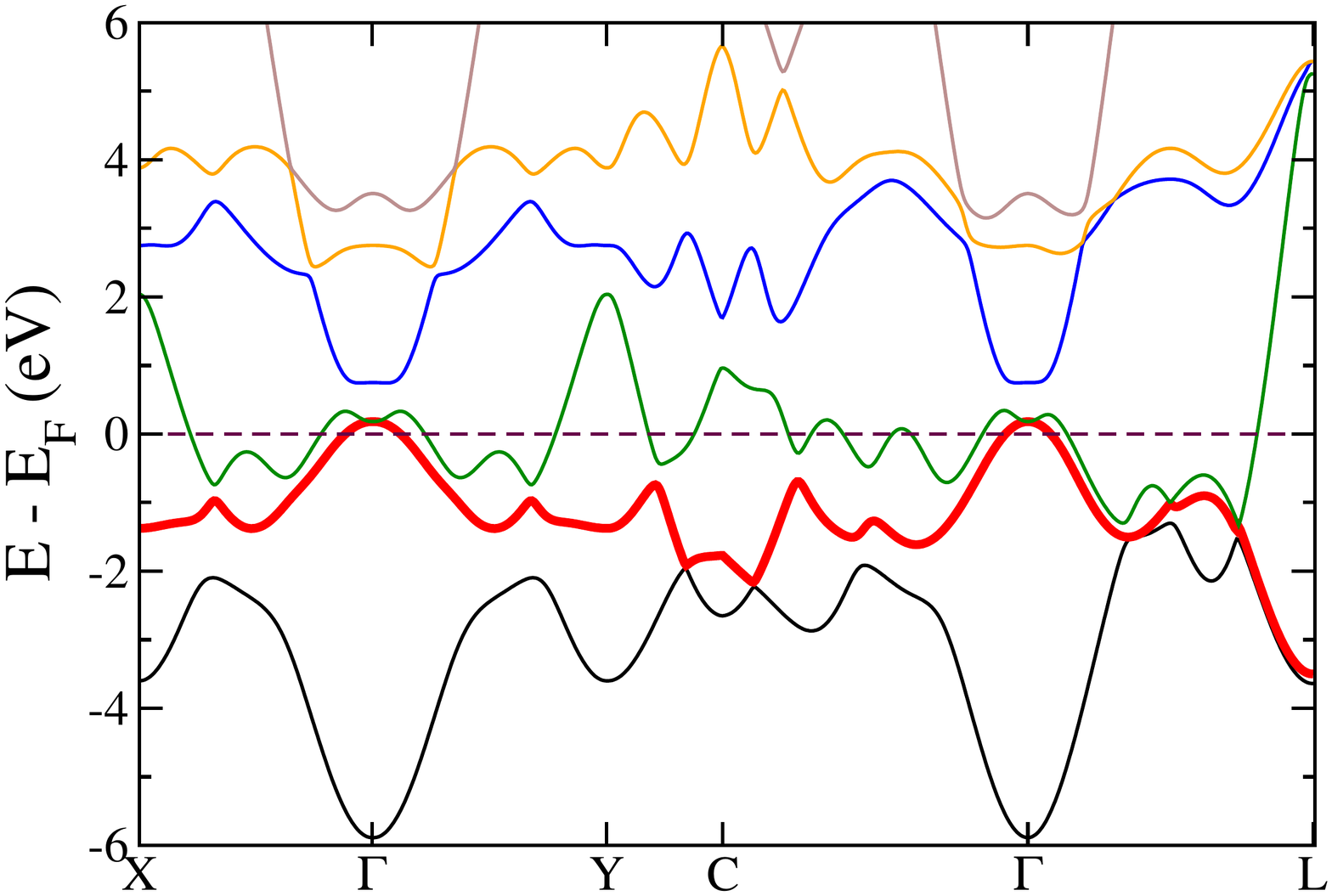}
    \includegraphics[width=0.49\textwidth]{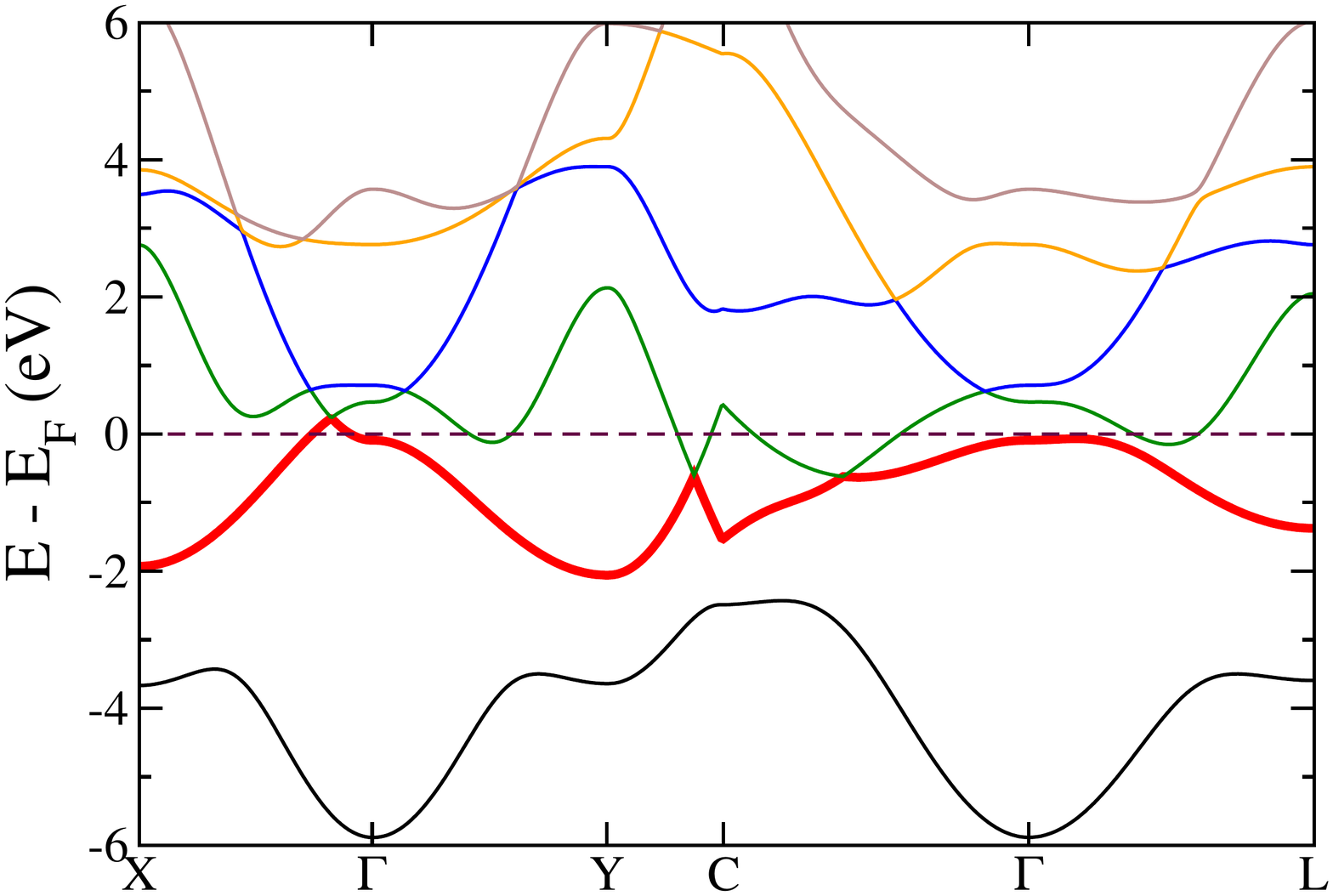}
    \caption{Band structure of tetragonal (left) and orthorhombic (right) Nb at 4.2\% uniaxial strain in a face-centered orthorhombic setting. The red band has the same $d_{yz}$ orbital character in both structures but different occupancies at the $\Gamma$ point, and is hence highlighted by a thicker line.}
    \label{fig:Nb-Ek-plots}
\end{figure}

\begin{figure}
	\centering
	\includegraphics[width=\textwidth]{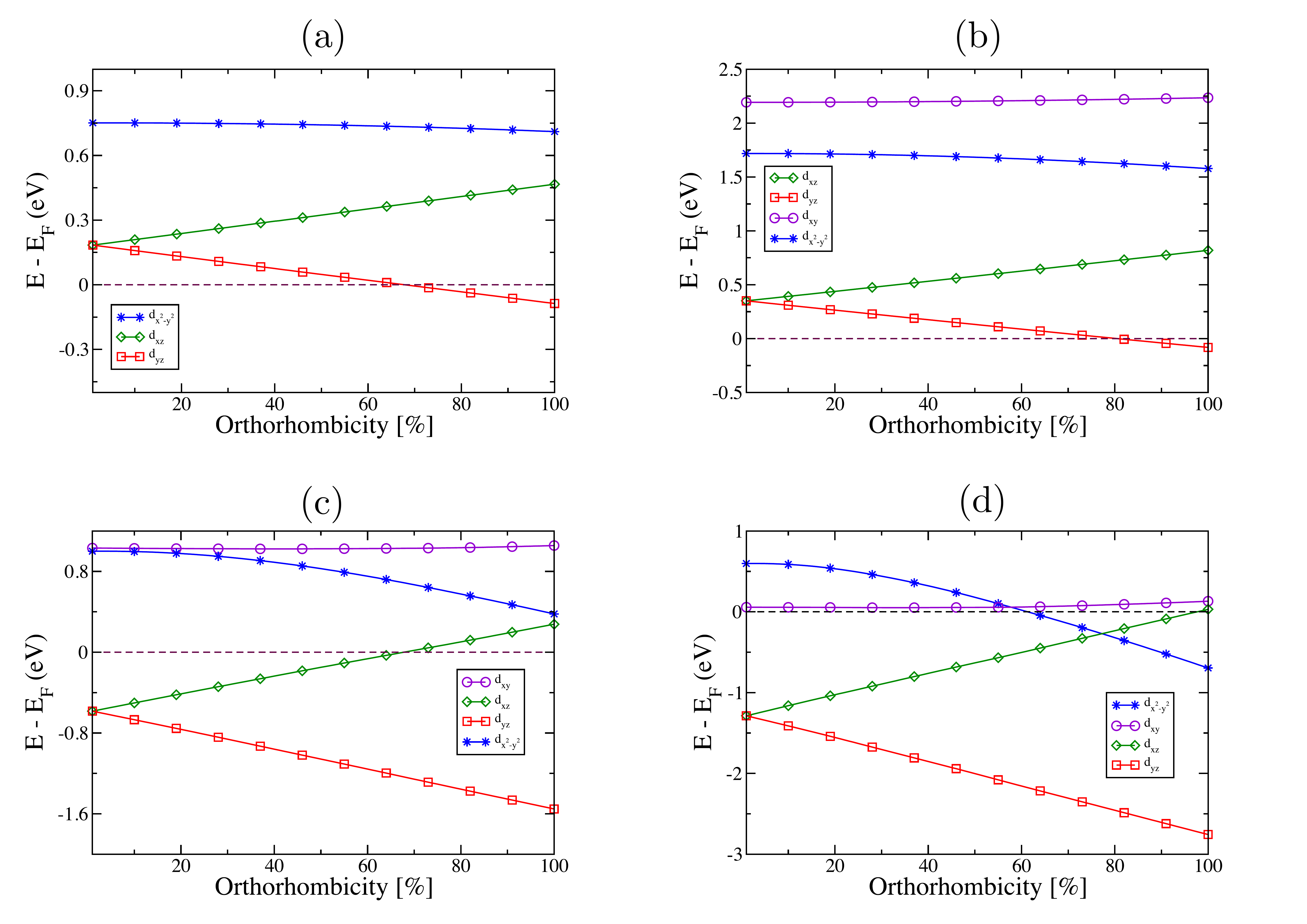}
	\caption{Band energies at $\Gamma$ point as a function of orthorhombicity for (a) BCC Nb at 4.2\% uniaxial strain (b) cP2 NbZr at 12\% uniaxial strain; (c)  cP2 MoNb at 16\% uniaxial strain; (d) BCC Mo at 18.1\% uniaxial strain.}
	\label{fig:NbZr-gamma-plot}
\end{figure}

Figure \ref{fig:Nb-Ek-plots} shows the band structure of tetragonal Nb and orthorhombic Nb at 4.2\% uniaxial strain along the $z$-direction. For consistency, we employ a face-centered orthorhombic primitive cell in both cases. Specifically, focus on the band represented by the thick red line. In the tetragonal case, the band lies above the Fermi energy at the $\Gamma$ point and is hence unoccupied. But in the orthorhombic case, this band is below the Fermi energy. This implies that the behavior of the red band at the $\Gamma$ point potentially plays a major role in the orthorhombic transition.

To gain further information about the band properties, we vary the extent of the orthorhombic symmetry breaking and study the corresponding variation of the band energies at the $\Gamma$ point. Let $a^{\prime}$ and $b^{\prime}$ refer to the equilibrium lattice constants of the orthorhombic structure in the $x$ and $y$ direction and let $a$ refer to the corresponding tetragonal lattice constant. We define ``orthorhombicity" as a linear interpolation of lattice parameters from tetragonal ($a$) to orthorhombic ($a^{\prime},b^{\prime}$).  Figure \ref{fig:NbZr-gamma-plot}a shows the band energies at the $\Gamma$ point, labeled according to orbital character, as a function of the orthorhombicity. Focus on the bands associated with $d_{xz}$ (green) and $d_{yz}$ (red) character. Initially, they are degenerate due to tetragonal symmetry and they are unoccupied because $E > E_{F}$. As the structure becomes more orthorhombic, the $d_{xz}$ band energy increases, moving further away from the Fermi energy. The $d_{yz}$ band energy decreases until it crosses the Fermi energy and becomes occupied. 

\begin{figure}
	\includegraphics[width=\textwidth]{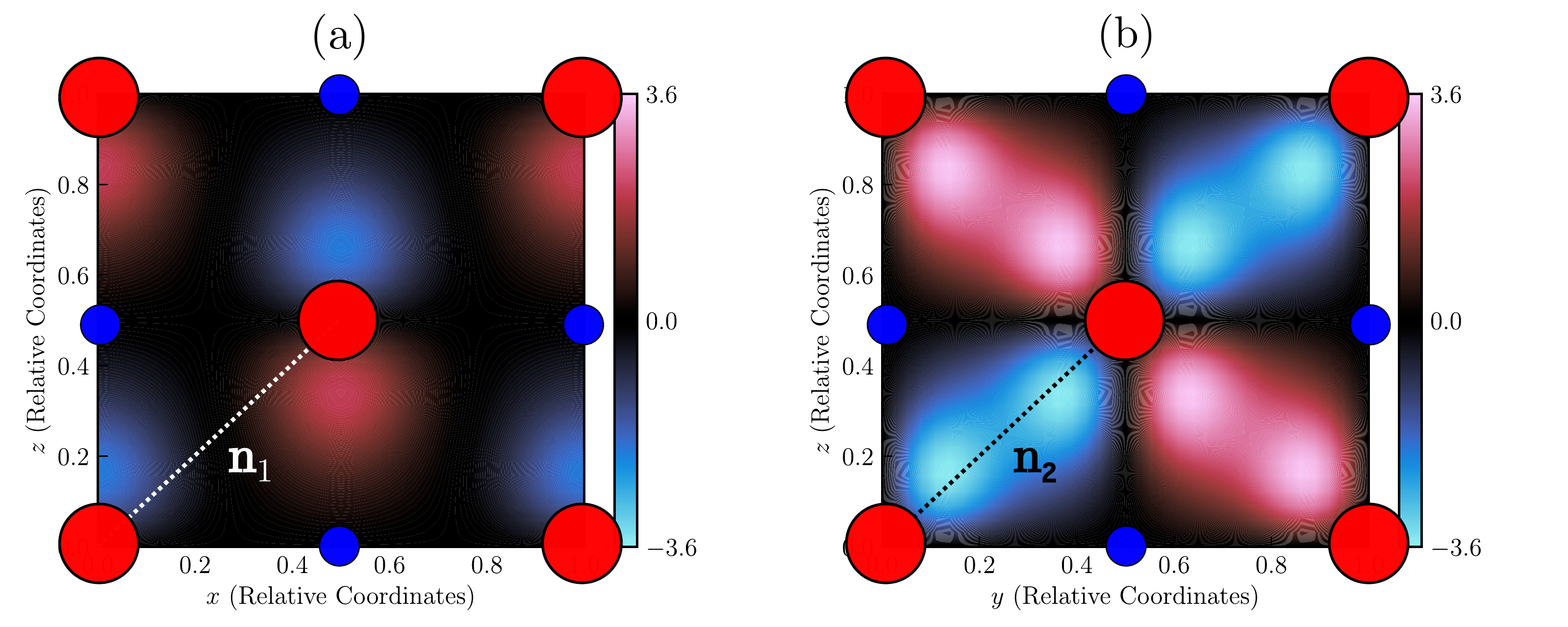}
	\caption{$d_{yz}$ wavefunction for 4.2\% strained orthorhombic Nb as seen from (a) $xz$ plane (anti-bonding); (b) $yz$ plane (bonding). Large and small atoms are at two different vertical heights. Colorbars denote the value of the $\Gamma$-point wavefunction.}
	\label{fig:Nb_dyz}
\end{figure}

To understand the consequence of the $d_{yz}$ band crossing $E_{F}$, we look at its wavefunction. The wavefunction is three dimensional, however we will look at 2D cuts in the $yz$ and $xz$ planes, as these planes contain the short and long near neighbor bonds respectively (Figure \ref{fig:Nb_dyz}). The atoms are depicted at the corresponding coordinates in the figures. The atoms in the plane of the wave function are drawn as red circles and the atoms that lie below the plane are drawn as blue circles. The near neighbor bond connects the atoms at (0,0) and (0.5,0.5). In the $yz$ projection, for which the near-neighbor bond  is the ``short bond" $\bm{n}_2$, the wavefunction maintains the same sign between the near neighbor atoms, which is indicative of bonding. Conversely, the wavefunction sign alternates between the near-neighbor atoms in the $xz$ projection (i.e the ``long bond" $\bm{n}_1$), which is indicative of anti-bonding. This explains the reason for the shortening of the $yz$ near-neighbor bonds (reducing $a$ to $b^{\prime}$) and the lengthening of the $xz$ near-neighbor bonds (increasing $a$ to $a^{\prime}$). Thus we understand that the orthorhombic distortion occurs to occupy the covalent bond that lowers the total energy.

\subsection{Universality}
We claim that the mechanism of preferentially occupying a short bonding orbital holds universally for Zr-Mo-Nb alloys over a range of valence electron count (VEC). Figure \ref{fig:NbZr-gamma-plot}b shows band energy at the $\Gamma$ point for cP2 NbZr at 12\% uniaxial strain as a function of orthorhombicity. The plot looks similar to the one seen for pure Nb, with the bonding $d_{yz}$ band crossing the Fermi energy and becoming occupied. The tetragonal $d_{yz}/d_{xz}$ degenerate energy for NbZr is further above $E_{F}$ than it was for cI2 Nb, due to the addition of Zr (valence 4), which lowers the VEC, hence lowering the Fermi energy. Further reduction in VEC would cause the BCC structure to lose local mechanical stability and transform to HCP {\em{via}} a Burgers distortion \cite{FengWidom2018}

We raise the VEC by alloying Nb with Mo (valence 6). Figure \ref{fig:NbZr-gamma-plot}c shows the band energy at the $\Gamma$ point for cP2 MoNb at 16\% uniaxial strain as a function of its orthorhombicity. Again, the $d_{xz}$ and $d_{yz}$ behave in a similar manner as seen for pure Nb, however in the tetragonal case for MoNb, the two bands are initially occupied. This is due to the increased VEC, which raises the Fermi energy. The $d_{xz}$ acts in opposition to the $d_{yz}$ band; For $d_{xz}$, the short direction ($yz$ plane) is anti-bonding and the long direction ($xz$ plane) is bonding. When both $d_{xz}$ and $d_{yz}$ are occupied, there is no overall net force driving orthorhombicity. As the orthorhombicity increases, the $d_{xz}$ band energy crosses the Fermi energy and becomes unoccupied. Now the occupied $d_{yz}$ orbital seeks to shorten the $y$-axis and lengthen the $x$-axis.

\begin{figure}
	\centering
	\includegraphics[width=\textwidth]{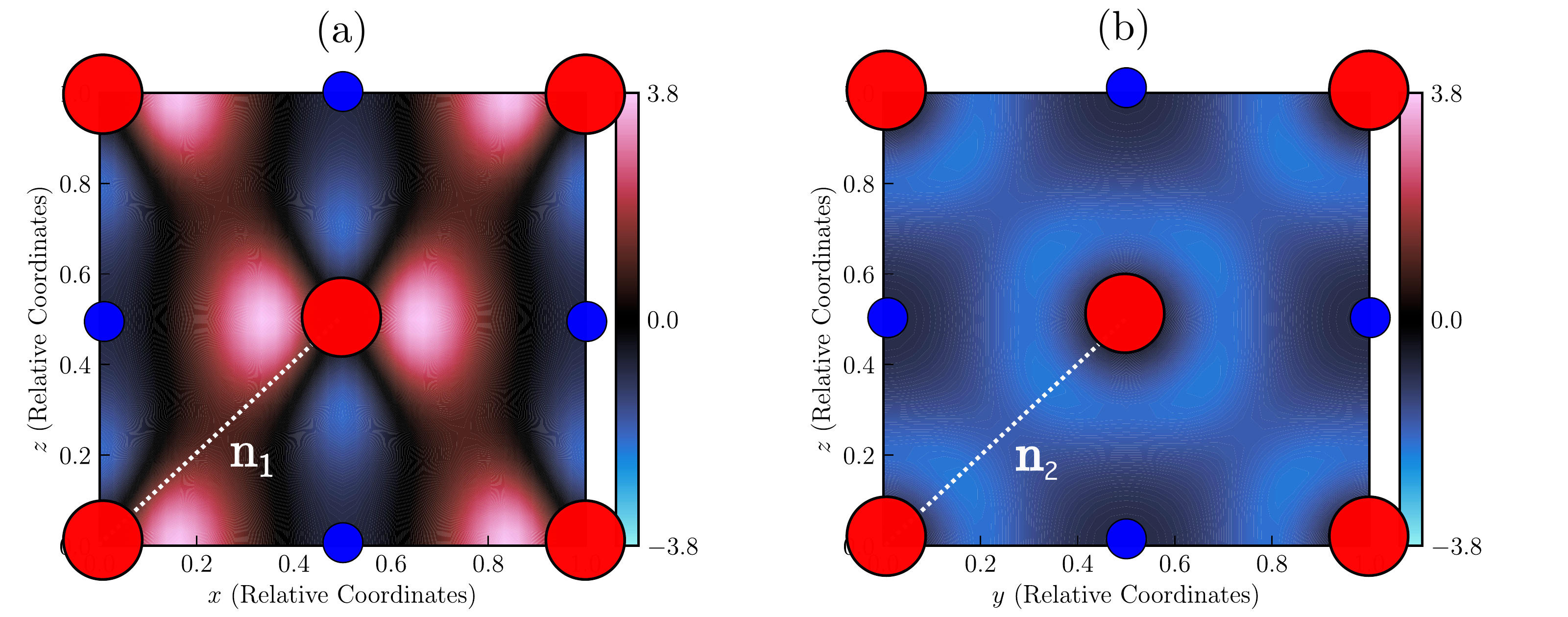}
	\caption{$d_{x^2 - y^2}$ wavefunction for orthorhombic Mo at 18.1\% uniaxial strain, as seen from (a) $xz$ plane (anti-bonding); (b) $yz$ plane (bonding). Large and small atoms are at two different vertical heights. Colorbars denote the value of the $\Gamma$-point wavefunction.}
	\label{fig:Mo-dx2y2-plot}
\end{figure}

Finally, consider the limit of pure Mo. Figure \ref{fig:NbZr-gamma-plot}d shows the band energy at the $\Gamma$ point at 18.1\% uniaxial strain as a function of orthorhombicity. Again $d_{xz}$ and $d_{yz}$ behave in a similar manner as seen for NbZr-MoNb, but the gap between the degenerate energy at the tetragonal limit and Fermi level is much larger. As a result, the $d_{xz}$  band is barely able to cross the Fermi energy, and another band ($d_{x^2-y^2}$)  drives the transition. Figure \ref{fig:Mo-dx2y2-plot} shows the $\Gamma$ point wavefunction for the $d_{x^2-y^2}$ band projected onto the $yz$ and the $xz$ planes. The $yz$ projection maintains the same sign between the near neighbor atoms, implying bonding, while the $xz$ projection shows a sign reversal, implying anti-bonding. As a result, the $yz$ near neighbor bond is shortened and the $xz$ bond is extended. 

In all cases, the transition occurs due to the presence of bonding and anti-bonding orbitals in the bands near the Fermi energy. For Nb, cP2 MoNb and cP2 NbZr, the degenerate pair $d_{xz}/d_{yz}$ drive the transition. While this pair is present in pure Mo, and behaves in a similar manner as seen in the other cases, it is the $d_{x^2-y^2}$ band that goes from unoccupied and occupied and is bonding along the $yz$ direction.

\section{Variation of Strain Thresholds}
\begin{figure}
	\includegraphics[width=0.5\textwidth]{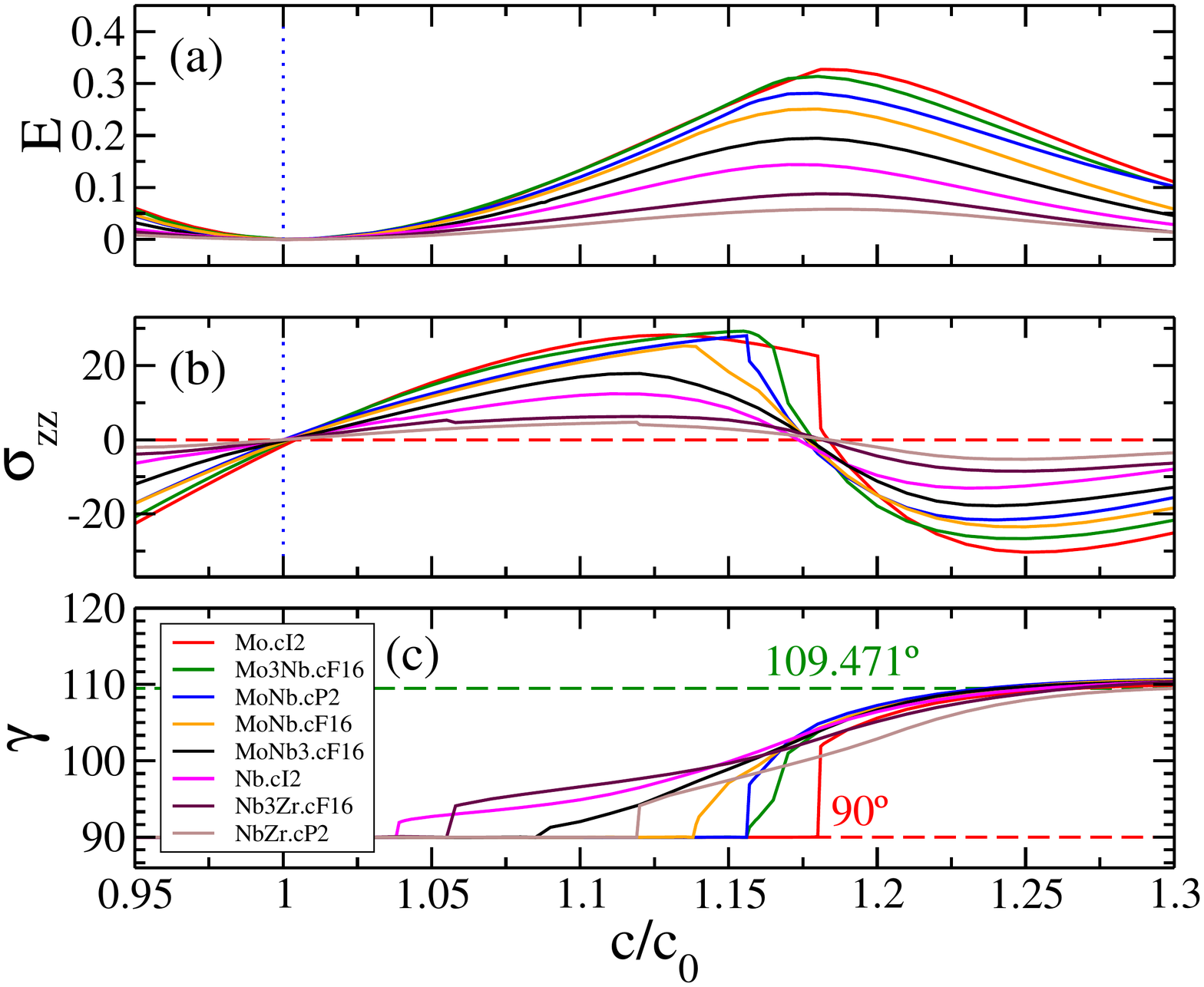}
	\caption{\label{fig:all-data}  Data for refractory metals and alloys under uniaxial strain with constant $c$ ($c_0$ is the equilibrium value). Graphs (a), (b), and (c), show energy $E$ (units eV/atom), stress $\sigma_{zz}$ (units GPa), and unit cell angle $\gamma$ (units $^\circ$), respectively. Valence electron count increases in steps of 0.25 from 4.5 for NbZr up to 6.0 for Mo.cI2.}
\end{figure}

\begin{table}
	\caption{\label{tab:vec} Variation with respect to VEC}
	\begin{tabular}{ll| rrrrrr | rr | rr}
		\hline
		Compound/ & Pearson/ & \multicolumn{6}{c|}{SOEC} & \multicolumn{2}{c|}{Splay} & \multicolumn{2}{c}{Extension} \\
		(VEC) & Group & $C_{11}$ & $C_{12}$ & $C_{44}$ & $C'$  & $\nu$ & $E$ & Strain&Angle & Strain&Barrier\\
		\hline
		NbZr (4.5) & cP2 ($Pm\bar{3}m$) & 150 & 111 & 17.8 & 19.5 & 0.424 & 56.3 & 12\% & 94.1$^{\circ}$ & 11.9\% & 0.039 \\
		Nb$_3$Zr (4.75) & cF16 ($Fm\bar{3}m$) & 195 & 122 & 14.7 & 36.5 & 0.385 & 101 & 5.8\% & 94.1$^\circ$ & 12\% & 0.056 \\
		Nb (5.00) & cI2 ($Im\bar{3}m$) & 247 & 137 & 16.2 & 55 & 0.357 & 149 & 3.9\% & 91.9$^\circ$ & 11.7\% & 0.098 \\
		MoNb$_3$ (5.25) & cF16 ($Fm\bar{3}m$) & 289 & 147 & 19.1 & 70.8 & 0.338 & 190 & 9.0\% & 91.3$^\circ$ & 12\% & 0.129 \\
		MoNb (5.50) & cF16 ($Fd\bar{3}m$) & 370 & 143 & 55.8 & 114 & 0.278 & 291 & 13.9\% & 91.7$^{\circ}$ & 13.5\% & 0.192\\
		MoNb (5.50) & cP2 ($Pm\bar{3}m$) & 379 & 140 & 63.8 & 119 & 0.270 & 302 & 15.7\% & 96.9$^\circ$ & 15.6\% & 0.259 \\
		Mo$_3$Nb (5.75) & cF16 ($Fm\bar{3}m$) & 431 & 145 & 83.5 & 143 & 0.252 & 358 & 15.7\% & 91.3$^\circ$ & 15.5\% & 0.275 \\
		Mo (6.00) & cI2 ($Im\bar{3}m$) & 467 & 160 & 99.4 & 154 & 0.255 & 386 & 18.1\% & 101.9$^\circ$ & 13\% & 0.209
	\end{tabular}
\end{table}

Figure~\ref{fig:all-data} illustrates the dependence of splay and extension thresholds on the VEC, by mixing Nb (valence 5) with Mo (valence 6) and Zr (valence 4) in varying proportions. Nominal values are given in Table \ref{tab:vec}. The threshold strain of the splay instability is the smallest at VEC = 5 (cI2 Nb). This is due to the separation of the $d_{xz}/d_{yz}$ band energy from the Fermi level in the tetragonal limit, which is smaller for Nb as compared to the other systems. At lower VEC , the gap increases while maintaining the same sign.  As VEC increases, the Fermi level rises, so that the $d_{xz}/d_{yz}$ band drops relative to $E_{F}$ and crosses in the vicinity of VEC = 5, leading to a minimum threshold for splay instability.

As seen in Table \ref{tab:vec}, the Young's modulus \cite{young-poisson}
\begin{equation}
	E = \frac{\sigma_{zz}}{u_{zz}} = \frac{(C_{11} - C_{12})(C_{11} + 2C_{12})}{C_{11} + C_{12}}\;[\rm{cubic}]
\end{equation}
for uniaxial strain along the $z$ direction increases with increasing VEC (here $\sigma_{zz}$ corresponds to the stress in the $zz$-direction). Hence, the slope of the stress-strain curve increases with VEC. Consequently, the maximum values of stress show a similar trend. Beyond their maxima, the systems become extensionally unstable. As demonstrated in Section \ref{sec:elastic}, this extensional instability for a tetragonal system occurs when the $w^{-}_{zz}$ eigenvalue vanishes. The extensional instabilities precede splay for NbZr, and for systems with VEC $\geq$ 5.5. The energies at the stress maxima correspond to barriers for the extensional instability. The energy maxima occur at strains higher than the stress maxima and correspond to the zero of the stress for all systems. Table \ref{tab:vec} also contains the Poisson ratio \cite{young-poisson}
\begin{equation}
	\nu = -\frac{u_{xx}}{u_{zz}} = \frac{C_{12}}{C_{11} + C_{12}}\;[\rm{cubic}]
\end{equation} 
for induced strain in the $x$/$y$ direction (due to applied strain along $z$). The Poisson ratio decreases with increasing VEC. Table \ref{tab:vec} also lists the elastic constants, splay angle, splay threshold, extension threshold and the barrier to the extensional instability for all the system studied. The elastic constants and the extension barrier largely increase along with the VEC, while the splay angle and extension threshold vary weakly. 
\section{Conclusion}
\label{sec:conc}
This paper provides a detailed description of the transitions in BCC refractory metals, triggered by the application of uniaxial strain. The structures initially become tetragonal, but after a certain strain threshold, the symmetry breaks, the cubic lattice vectors splay, and the structure becomes orthorhombic. On the basis of first-principles energies, we show that this transition is discontinuous, marked by an overlap region where both tetragonal and orthorhombic solutions are locally mechanically stable, but global stability interchanges as shown in Figure \ref{fig:Mo-17.5}. A simple nonlinear Landau-type expansion (Appendix \ref{app:B}) models the transition and explains it's discontinuous nature. 

The geometry of orthorhombic structures was studied using first-principles relaxation. The symmetry breaking that creates ``short" and ``long" near-neighbor bonds indicates that the orthorhombic transition is a type of Jahn-Teller-Peierls distortion. First-principles electronic band structures and wavefunctions provide insight into the transition mechanism.  Specifically, as shown in Figure \ref{fig:NbZr-gamma-plot}, the degeneracy between the bands of $d_{xz}/d_{yz}$ splits as the extent of orthorhombicity is increased. If the $yz$ plane contains the short bond, the $d_{yz}$ bonding orbital energy decreases while the $d_{xz}$ anti-bonding orbital energy increases. In most cases, one of these bands cross the Fermi energy. This results in the occupation of bonding orbitals along the short bond direction. At high VEC (e.g in the case of Mo), the $d_{x^2-y^2}$ band drives the transition, and it is also bonding along the short direction.  The $\Gamma$-point band energy plots also provide some insight into Wallace tensor behavior. The energy of the $d_{xz}/d_{yz}$ degenerate bands at the tetragonal limit varies strongly with VEC. For Nb, it is close to the Fermi energy, while for Mo, the energy gap is significantly larger. The implication is that the dependence of band energy on orthorhombicity affects the ductility of the system. Note that ductility is a complex phenomenon, and the orthorhombicity is only one of many factors whose interplay determines the ductility of the system.

We also look at the eigenvalues of the symmetric Wallace tensor. For Nb, the transition prevented one of the shear eigenvalues from going negative, keeping the system mechanically stable. The orthorhombic transition hence explains the high intrinsic ductility parameter of Nb. Conversely for Mo, the system undergoes an extensional instability before reaching the shear threshold. These results are consistent with previous theoretical work \cite{qi-chrzan,dejong,winter,curtin}.

Finally, we study the variation of the splay and extension thresholds, extension barrier, splay angle and elastic constants with the valence electron count (see Figure \ref{fig:all-data} and Table \ref{tab:vec}). The elastic constants, splay threshold and extension barrier display a strong variation with the VEC. However, the extension threshold and splay angle do not. The extensional instability occurs when the stress reaches a maximum and decreases on application of further strain. Equivalently, for tetragonal systems, the extensional instability is brought about when the $w^{-}_{zz}$ eigenvalue goes negative. Setting $w^{-}_{zz} = 0$ and using the analytical expression for $w^{-}_{zz}$ (given in Appendix \ref{app:A}) gives a relation between the stress and elastic constants at the extensional threshold. However, the physical mechanism driving the extensional instability is unknown and further study is needed to gain a deeper understanding into its origin and behavior.

The ability to tune the splay threshold can be useful for refractory HEA design. The average VEC of the alloy will significantly affect the onset of the splay transition. At a microstructure level, the near neighbor bonding plays a major role in determining the splay threshold. A random MoNb alloy at fixed VEC may show differing splay properties depending on the number and distribution of Mo-Mo, Mo-Nb and Nb-Nb near neighbor bonds. The shear modulus, Young's modulus and Poisson ratio also provide an indication on which instability (splay or extensional) occurs at lower strain. Carefully tuning these parameters enables control over the ductility. Point defects, dislocations and other microstructure effects will also affect the ductility of the system \cite{curtin,otherfactors1,otherfactors2,otherfactors3,otherfactors4,otherfactors5}, but are out of the scope of this work.

\begin{acknowledgments}
This work was supported by the ARPA-E program at CMU (grant DE-AR0001430) and at NETL. Investigation of the role of electronic band structure was supported by the Department of Energy grant DE-SC0014506. This work was funded by the Department of Energy, ARPA-e, an agency of the United States Government. Neither the United States Government, nor Carnegie Mellon University, nor any agency thereof, nor any of their employees, makes any warranty, expressed or implied, or assumes any legal liability or responsibility for the accuracy, completeness, or usefulness of any information, apparatus, product, or process disclosed, or represents that its use would not infringe privately owned rights. Reference herein to any specific commercial product, process, or service by trade name, trademark, manufacturer, or otherwise, does not necessarily constitute or imply its endorsement, recommendation, or favoring by the United States Government or any agency thereof. The views and opinions of authors expressed herein do not necessarily state or reflect those of the United States Government or any agency thereof.
\end{acknowledgments}

\begin{appendix}

\section{Calculation of Wallace tensor eigenvalues and eigenvectors}
\label{app:A}
\subsection{Tetragonal symmetry}
As shown previously in Section \ref{sec:Methods}, the symmetrized Wallace tensor (SWT) for the tetragonal structure in Voigt notation is
\begin{equation}
    W_{ij} = \begin{pmatrix}
    C^{\prime}_{11} & C^{\prime}_{12} & C^{\prime}_{13} - \frac{\tau}{2} & 0 & 0 & 0 \\
    C^{\prime}_{12} & C^{\prime}_{11} & C^{\prime}_{13} - \frac{\tau}{2} & 0 & 0 & 0 \\
    C^{\prime}_{13} - \frac{\tau}{2} & C^{\prime}_{13} - \frac{\tau}{2} & C^{\prime}_{33} + \tau & 0 & 0 & 0 \\
    0 & 0 & 0 & C^{\prime}_{44} + \frac{\tau}{2} & 0 & 0 \\
    0 & 0 & 0 & 0 & C^{\prime}_{44} + \frac{\tau}{2} & 0 \\
    0 & 0 & 0 & 0 & 0 & C^{\prime}_{66}
    \end{pmatrix},
\end{equation}
where $C^{\prime}$ is the second order elastic tensor at finite strain and $\tau$ is the second Piola-Kirchoff (PK2) stress. Three of the eigenvalues $w$ (and eigenvectors $v$) can be obtained trivially by observing the lower right part of the matrix
\begin{align}
    w_{xy}v_{xy} = C^{\prime}_{66}\begin{pmatrix}
    0 \\
    0 \\
    0 \\
    0 \\
    0 \\
    1 
    \end{pmatrix}, w_{yz}v_{yz} = \left(C^{\prime}_{44} + \frac{\tau}{2}\right)\begin{pmatrix}
    0 \\
    0 \\
    0 \\
    1 \\
    0 \\
    0 
    \end{pmatrix}, w_{xz}v_{xz} = \left(C^{\prime}_{44} + \frac{\tau}{2}\right)\begin{pmatrix}
    0 \\
    0 \\
    0 \\
    0 \\
    1 \\
    0 
    \end{pmatrix}.
\end{align}
This reduces the problem significantly. The $v_{yz}$ and $v_{xz}$ shear modes are degenerate. This is due to the symmetry equivalency between the $xz$ and $yz$ directions in the tetragonal crystal structure. 

We now need to solve the eigenvalues and eigenvectors of the top left 3 $\times$ 3 matrix. Let's define $D_{11}  = C^{\prime}_{11},\;D_{12} = C^{\prime}_{12},\;D_{13} = C^{\prime}_{13} - \frac{\tau}{2}$ and $D_{33} = C^{\prime}_{33} + \tau$. We write the characteristic polynomial in terms of these new parameters as
\begin{equation}
    \begin{vmatrix}
    \lambda - D_{11} & -D_{12} & -D_{13} \\
    -D_{12} & \lambda - D_{11} & -D_{13} \\
    -D_{13} & -D_{13} & \lambda - D_{33}
    \end{vmatrix} = 0,
\end{equation}
Solving for the roots of the resulting cubic polynomial is simplified by observing that
\begin{equation}
    v_{xxyy} = \frac{1}{\sqrt{2}}\begin{pmatrix}
    1 \\
    -1 \\
    0 \\
    0 \\
    0 \\
    0
    \end{pmatrix}
\end{equation}
is an eigenvector with the eigenvalue $w_{xxyy} = D_{11} - D_{12} = C^{\prime}_{11} - C^{\prime}_{12}$, providing one of the roots of the cubic polynomial. We then use relations between polynomial roots and coefficients to obtain 
\begin{equation}
    w^{\pm}_{zz} = \frac{D_{11} + D_{12} + D_{33} \pm \sqrt{(D_{11} - D_{12} - D_{33})^{2} + 8D^{2}_{13}}}{2}.
\end{equation}
In terms of elastic constants and stress, we obtain
\begin{equation}
    w^{\pm}_{zz}= \frac{C^{\prime}_{11} + C^{\prime}_{12} + C^{\prime}_{33} + \tau \pm \sqrt{(C^{\prime}_{11} - C^{\prime}_{12} - C^{\prime}_{33} - \tau)^{2} + 8\left(C^{\prime}_{13} - \frac{\tau}{2}\right)^{2}}}{2}.
\end{equation}
The minus solution is smaller, and will be the eigenvalue that eventually triggers an extensional instability.

Symmetry provides a simple description for the form of the extensional eigenvectors. When a strain is applied along the $z$-direction, we expect an induced Poisson strain along the $x$ and $y$ direction. Now in a tetragonal crystal, when viewing along [001], the $x$ and $y$ directions are indistinguishable. To rephrase in Voigt terms, the $xz$ and $yz$ directions are symmetric. As a result, the induced strains along $x$ and $y$ should be identical. This implies that the eigenvectors associated with the extensional modes should have the form
\begin{equation}
	v^{\pm}_{zz} \equiv \begin{pmatrix}
		a \\
		a \\
		b \\
		0 \\
		0 \\
		0 \\
	\end{pmatrix}.
\end{equation}
Employing the eigenvalue equations $Wv^{\pm}_{zz} = w^{\pm}_{zz}v^{\pm}_{zz}$, we write
\begin{align}
	v^{+}_{zz} &= \frac{1}{\sqrt{1 + 2(a^{+})^2}}\begin{pmatrix}
		a^{+}\\
		a^{+} \\
		1 \\
		0 \\
		0 \\
		0 \\
		\end{pmatrix}, v^{-}_{zz} = \frac{1}{\sqrt{1 + 2(a^{-})^2}}\begin{pmatrix}
		a^{-}\\
		a^{-}\\
		1 \\
		0 \\
		0 \\
		0 \\
	\end{pmatrix}, \\
a^{+} &= \frac{w^{+}_{zz} - C^{\prime}_{33} - \tau}{2\left(C^{\prime}_{13} - \frac{\tau}{2}\right)},\;a^{-} = \frac{w^{-}_{zz} - C^{\prime}_{33} - \tau}{2\left(C^{\prime}_{13} - \frac{\tau}{2}\right)}.
\end{align}
At small strain, 
\begin{equation}
	w^{+}_{zz} > C^{\prime}_{33} + \tau,\;w^{-}_{zz} < C^{\prime}_{33} + \tau.
\end{equation}
The induced Poisson strain for $v^{+}_{zz}$ is positive and for $v^{-}_{zz}$ is negative in the limit of small strain.
\subsection{Orthorhombic symmetry}
As shown previously in Section \ref{sec:Methods}, the SWT for the orthorhombic structure in Voigt notation is
\begin{equation}
	W_{ij} = \begin{pmatrix}
		C^{\prime}_{11} & C^{\prime}_{12} & C^{\prime}_{13} - \frac{\tau}{2} & 0 & 0 & 0 \\
		C^{\prime}_{12} & C^{\prime}_{22} & C^{\prime}_{23} - \frac{\tau}{2} & 0 & 0 & 0\\
		C^{\prime}_{13} - \frac{\tau}{2} & C^{\prime}_{23} - \frac{\tau}{2} & C^{\prime}_{33} + \tau & 0 & 0 & 0 \\
		0 & 0 & 0 & C^{\prime}_{44} + \frac{\tau}{2} & 0 & 0 \\
		0 & 0 & 0 & 0 & C^{\prime}_{55} + \frac{\tau}{2} & 0 \\
		0 & 0 & 0 & 0 & 0 & C^{\prime}_{66}
	\end{pmatrix}.
\end{equation}
Three of the eigenvalues (and eigenvectors) are immediately recognizable
\begin{equation}
	w_{xy}v_{xy} = C^{\prime}_{66}\begin{pmatrix}
		0 \\
		0 \\
		0 \\
		0 \\
		0 \\
		1
	\end{pmatrix}\;,\;w_{yz}v_{yz} = (C^{\prime}_{44} + \frac{\tau}{2})\begin{pmatrix}
	0 \\
	0 \\
	0 \\
	1 \\
	0 \\
	0 
\end{pmatrix}\;,\;w_{xz}v_{xz} = (C^{\prime}_{55} + \frac{\tau}{2})\begin{pmatrix}
0 \\
0 \\
0 \\
0 \\
1 \\
0
\end{pmatrix}.
\end{equation}
Note that the degeneracy between $v_{yz}$ and $v_{xz}$ eigenvectors has been broken. To get the other three eigenvectors, the top-left 3x3 block has to be solved. In the tetragonal case, symmetry yielded a fourth eigenvector which simplified the calculations significantly. However, since $C^{\prime}_{11} \neq C^{\prime}_{22}$ for the orthorhombic system, the resulting cubic polynomial has to be solved numerically or symbolically. However, the three remaining eigenvectors will be a shear-extension mix. To see this, write the general form of the remaining eigenvectors
\begin{equation}
	v_{4,5,6} = \begin{pmatrix}
		\alpha \\ \beta \\ \gamma \\ 0 \\ 0 \\ 0 
		\end{pmatrix} = c_1\begin{pmatrix}
			1 \\ -1 \\ 0 \\ 0 \\ 0 \\ 0
		\end{pmatrix} + c_2\begin{pmatrix}
			a^{+} \\ a^{+} \\ 1 \\ 0 \\ 0 \\ 0
		\end{pmatrix} + c_3\begin{pmatrix}
		a^{-} \\ a^{-} \\ 1 \\ 0 \\ 0 \\ 0
	\end{pmatrix}
\end{equation}
The first basis vector is a tetragonal shear mode and the second and third vectors are tetragonal extensional modes. This means that the orthorhombic system will have 3 shear modes and 3 mixed modes. 
\section{Theoretical modeling of the discontinuous orthorhombic transition}
\label{app:B}
Start with a BCC system with initial lattice constant $a_0$ stretched along the $z$-direction by engineering strain $\zeta$. Applying the Poisson strain to the $x$ and $y$ components of the lattice vectors, the new lattice vectors of the system become
\begin{equation}
    \bm{L}_{\rm{tetra}} = \begin{bmatrix}
    a & 0 & 0 \\
    0 & a & 0 \\
    0 & 0 & c
    \end{bmatrix}.
    \label{eq:tetra}
\end{equation}
A deformation that breaks the tetragonal symmetry results in
\begin{equation}
    \bm{L}_{\rm{ortho}} = \begin{bmatrix}
    a^{\prime} & 0 & 0 \\
    0 & b^{\prime} & 0 \\
    0 & 0 & c
    \end{bmatrix}
    \label{eq:ortho}
\end{equation}
through a Green-Lagrange strain of the form
\begin{equation}
    \bm{\eta} = \begin{bmatrix}
    \alpha & -\alpha & 0 & 0 & 0 & 0
    \end{bmatrix}
    \label{eq:alpha-strain}.
\end{equation}
The energy (per volume) required to make such a deformation can be expressed as a Taylor series
\begin{equation}
    f(\bm{\eta}) = E_{\rm{ortho}} - E_{\rm{tetra}} =  \frac{1}{2!}C_{ij}\eta_i\eta_j + \frac{1}{4!}C_{ijkl}\eta_i\eta_j\eta_k\eta_l + \frac{1}{6!} C_{ijklmn}\eta_i\eta_j\eta_k\eta_l\eta_m\eta_n + \cdots,
\end{equation}
where $C_{ij}$, $C_{ijkl}$ and $C_{ijklmn}$ are the second, fourth and sixth order elastic constants of the tetragonal structure. We can simplify this expression significantly by using the symmetry of the elastic constants. In the tetragonal system, the $x$ and $y$ directions are indistinguishable. As a result, switching the $1$ indices with the $2$ indices in the elastic constants has no effect. We ignore terms beyond sixth order and substitute \ref{eq:alpha-strain} to get
\begin{align}
    f(\alpha) &= A\alpha^2 + B\alpha^4 + C\alpha^6, \\
    A &= C_{11} - C_{12}, \\
    B &= \frac{1}{12}\left(C_{1111} - 4C_{1112} + 3C_{1122}\right), \\
    C &= \frac{1}{360}\left(C_{111111} - 6C_{111112} + 15C_{111122} - 10C_{111222}\right).
    \label{eq:model-expression}
\end{align}
Solutions to this model can be found by solving for roots of the first derivative
\begin{equation}
    \frac{df}{d\alpha} = 2A\alpha + 4B\alpha^3 + 6C\alpha^5 = 0.
    \label{eq:first-deriv}
\end{equation}
The stability of the $\alpha = 0$ solution can be obtained by second derivative
\begin{equation}
    \frac{d^2 f}{d\alpha^2} = 2A + 12B\alpha^2 + 30C\alpha^4 = 0. 
\end{equation}
Setting $\alpha = 0$, $d^2f/d\alpha^2 = 2A$.  If $A > 0$ i.e $C_{11} > C_{12}$, the $\alpha = 0$ (tetragonal) solution is an energy minimum. If $C_{11} < C_{12}$, then the tetragonal solution is unstable. Dividing Eq. \ref{eq:first-deriv} by $2\alpha$ yields 
yields a quadratic in $\alpha^2$, and it's solutions can be written as
\begin{equation}
    \alpha^2 = \frac{1}{3C}\left(-B \pm \sqrt{B^2 - 3AC}\right),
    \label{eq:quad-solution}
\end{equation}
with stability $d^2f/d\alpha^2 = -8A - 8B\alpha^2$. If $B^2 > 3AC$ and $-B \pm \sqrt{B^2 - 3AC} > 0$, then two orthorhombic ($\alpha \neq 0$) solutions exist. If any of those two non-zero $\alpha$ solutions satisfy $-8A - 8B\alpha^2 > 0$, then this orthorhombic solution is an energy minimum. Furthermore, if in addition $A > 0$, then both the tetragonal and the orthorhombic solutions will be locally stable, resulting in coexistence and discontinuous transitions, matching what we observe in the contour plots.

\begin{figure}
    \centering
    \includegraphics[width=0.5\textwidth]{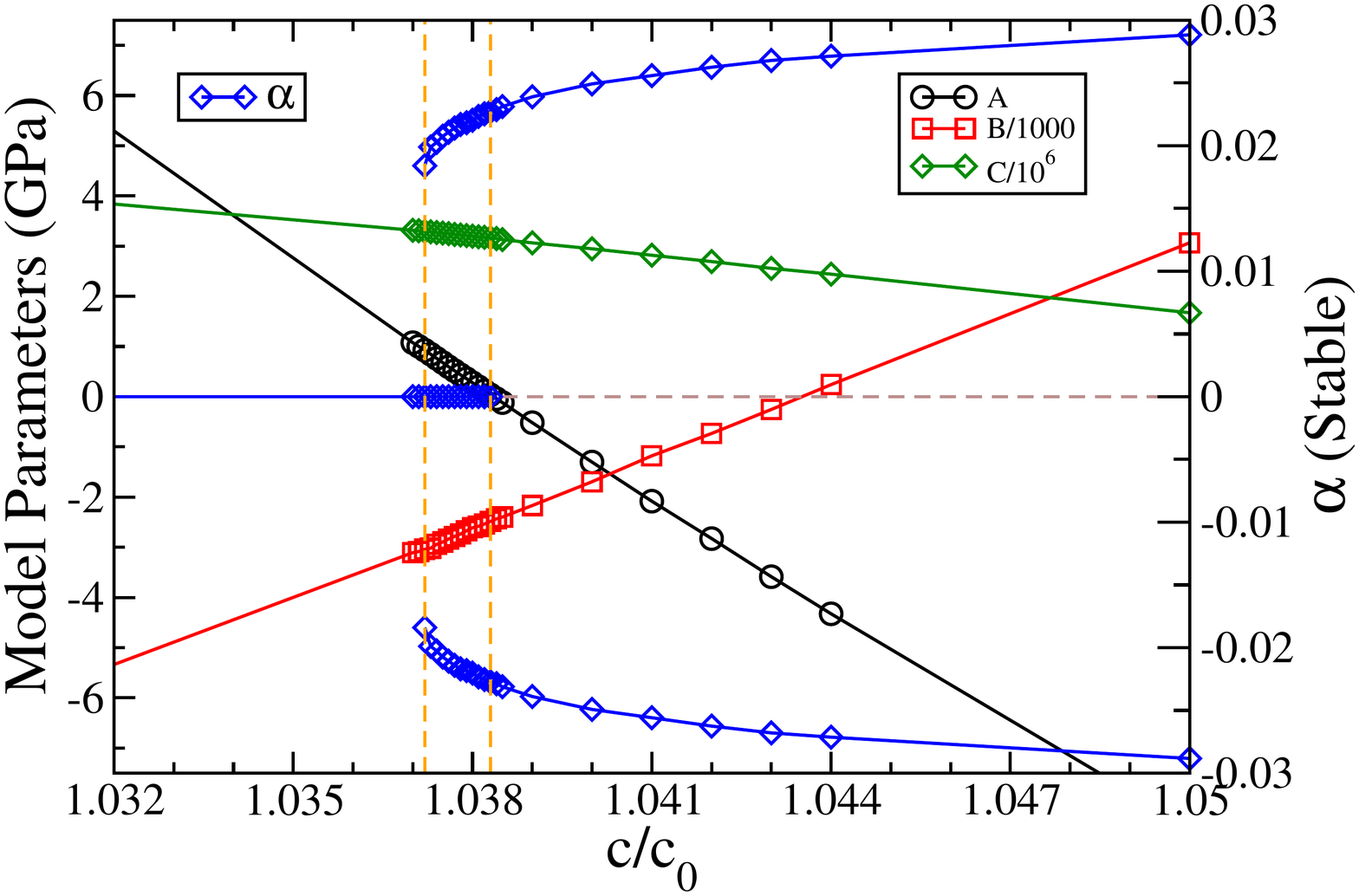}
    \caption{The evolution of the model parameters A, B and C, and the tetragonal symmetry breaking parameter $\alpha$ (denoted by the blue line) in the vicinity of the transition for Nb. The dashed orange lines mark the co-existence region.}
    \label{fig:Nb_ABC}
\end{figure}

Our model accurately describes the orthorhombic transition of Nb. Figure \ref{fig:Nb_ABC} shows the values of $A$, $B$ and $C$ as a function of the applied uniaxial strain. These parameters were evaluated by calculating the second, fourth and sixth order elastic constants for different uniaxial strains and employing (\ref{eq:model-expression}). It can be seen that $A$ is positive up to 3.8\% strain. The fourth order coefficient $B$ is negative in the vicinity of this strain, which allows for $\alpha \neq 0$ solutions even when $A > 0$. The sixth order coefficient $C$ is positive in this region. This ensures that an uncontrolled divergence of $\alpha$ is prevented. 

\begin{figure}
    \centering
    \includegraphics[width=0.5\textwidth]{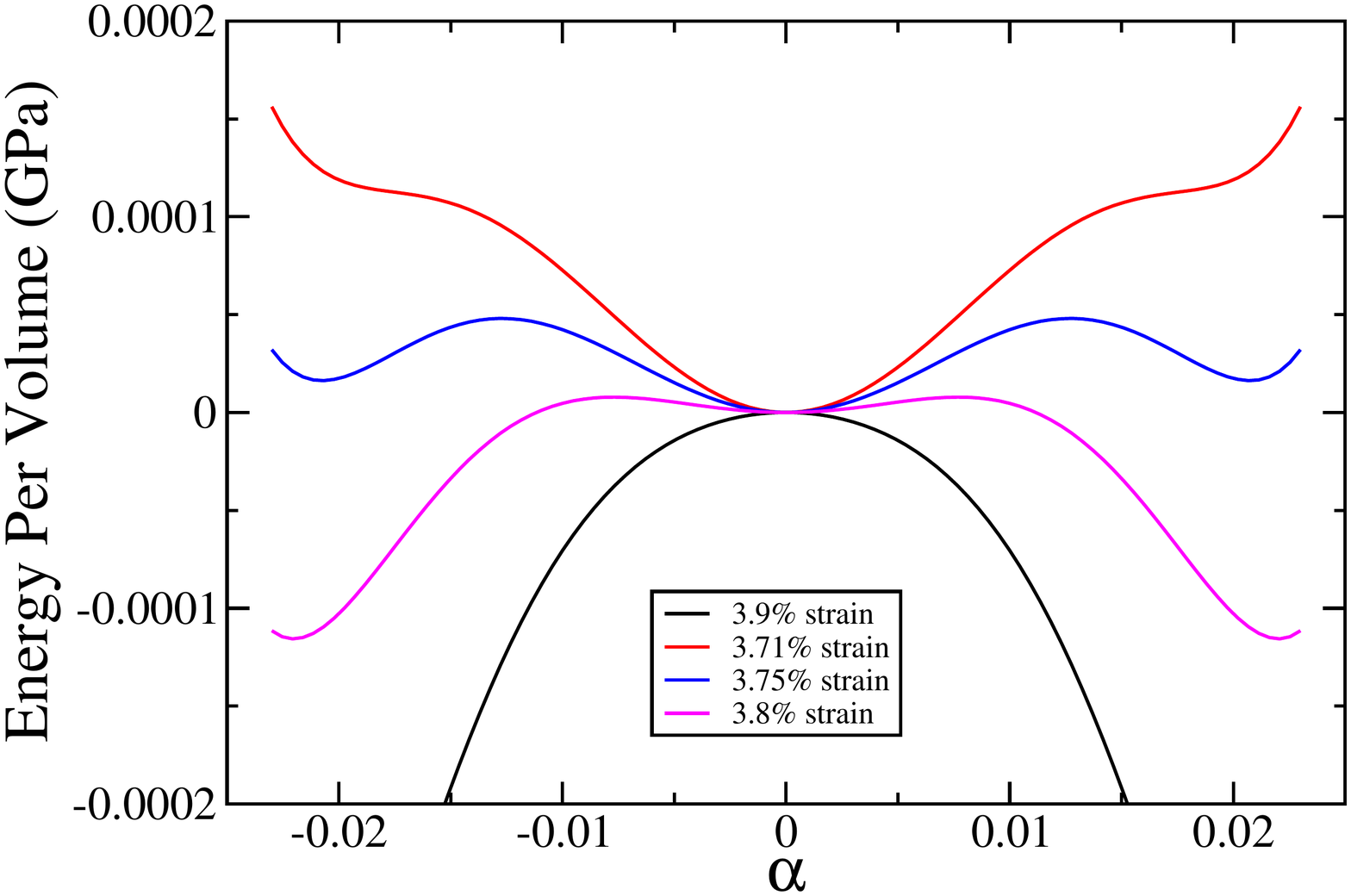}
    \caption{Model energy per volume (in GPa) vs $\alpha$, at multiple strains between 3.7-3.9\%. At 3.71\% strain, tetragonal is the only stable solution. At 3.9\%, orthorhombic is the only stable solution.}
    \label{fig:Nb_multi_solution}
\end{figure}

Plugging these parameters into (\ref{eq:quad-solution}) to calculate $\alpha$ for Nb yields $\alpha(\zeta)$. For strains up to 3.7\%, $B^2 - 3AC < 0$, and $\alpha = 0$ is the only real and stable solution. For 3.9\% strain and above, $\alpha = 0$ is no longer a stable solution and the system undergoes an orthorhombic deformation. In the interval 3.7-3.9\% strain both the tetragonal and orthorhombic solutions exist and are locally stable. To see this, we plot the model function (Equation  (\ref{eq:model-expression})) using $A$, $B$ and $C$ calculated at strains in this interval for different values of $\alpha$. This is shown in Figure \ref{fig:Nb_multi_solution}, where we can clearly see the two solutions. Note that the minima at positive and negative $\alpha$ correspond to the same solution, since $f(\alpha) = f(-\alpha)$.

\begin{figure}
    \centering
    \includegraphics[width=0.5\textwidth]{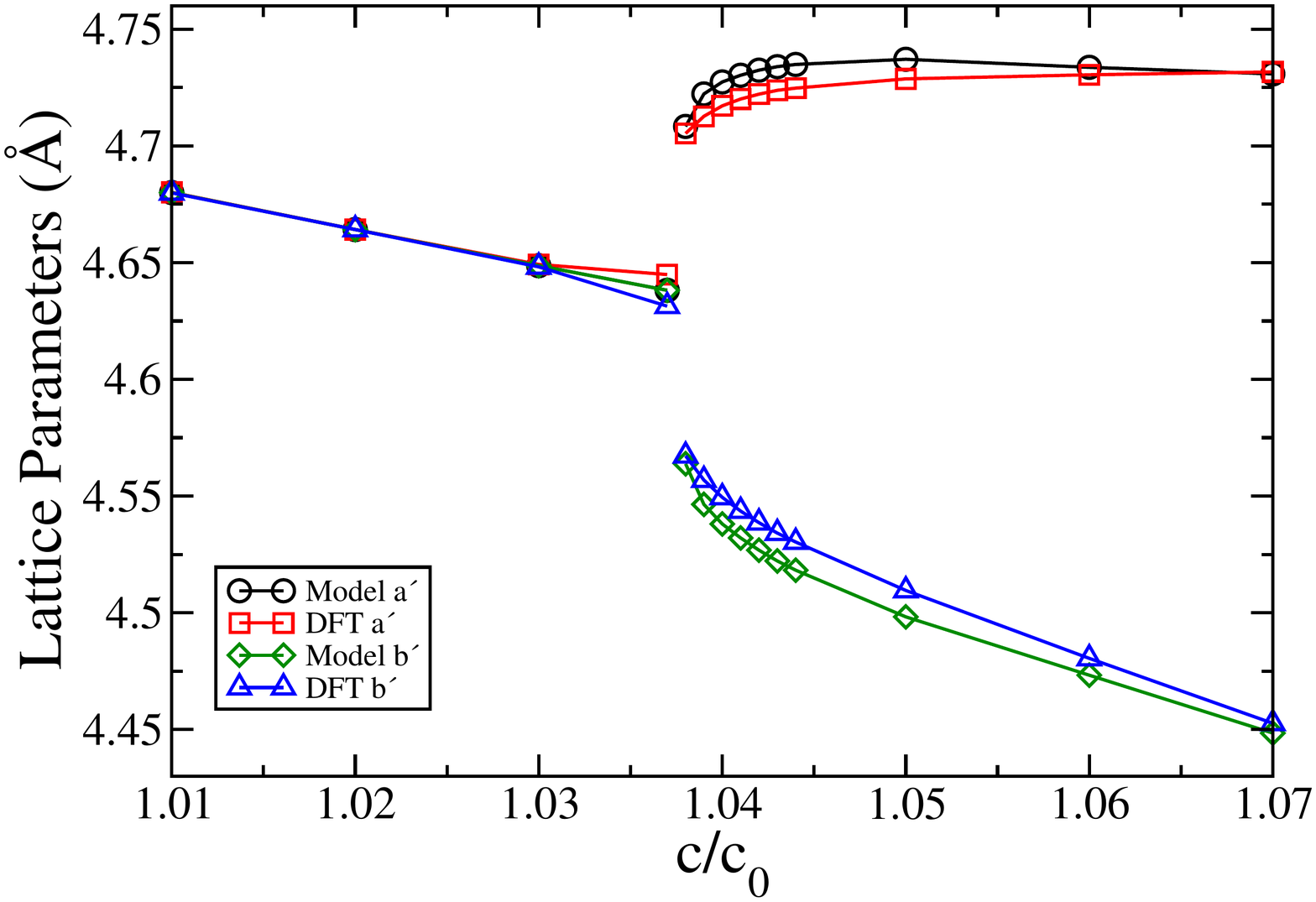}
    \caption{The evolution of the orthorhombic lattice parameters $a^{\prime}$ (black and red curves) and $b^{\prime}$ (blue and green curves) with uniaxial strain. The black and green curves were obtained from the model calculations, while the red and blue curves were obtained from first principles DFT. At 3.7\% strain and below, only tetragonal solution exists.}
    \label{fig:Nb_lp}
\end{figure}

Finally, we compare the orthorhombic lattice constants predicted by the model to those obtained from DFT. For a given $\alpha$, the orthorhombic lattice parameters can be written as
\begin{equation}
    a^{\prime} = a\sqrt{1 + 2\alpha}\;,\;b^{\prime} = a\sqrt{1 - 2\alpha}, 
\end{equation}
where $a^{\prime},b^{\prime},a$ are defined in (\ref{eq:tetra}) and (\ref{eq:ortho}). Figure \ref{fig:Nb_lp} shows the evolution of $a^{\prime}$ and $b^{\prime}$ with uniaxial strain. Model values compare very well to the first-principles values which helps validate the model.

In conclusion, we have obtained a simple model which describes the tetragonal to orthorhombic transition as an energy minimizing deformation. Importantly, it captures the discontinuous nature of transition, attributed mainly to the ability of the second and the fourth order coefficients to become negative. This is in contrast to the continuous transition model where only the second order term goes negative while the quartic remains positive. We provide additional validation for our model by comparing the predicted orthorhombic lattice constants with the DFT obtained values. 
\end{appendix}

\end{document}